\documentclass[
  journal=pasa,
  manuscript=research-paper,
  year=2025,
  volume=xx,
]{cup-journal}

\usepackage{graphicx}	
\usepackage{amsmath}	
\usepackage{amssymb}	
\usepackage{bm}
\usepackage[tracking=smallcaps]{microtype}
\SetTracking[no ligatures = f]{encoding = *,shape = sc}{0}
\usepackage{booktabs}

\usepackage{natbib} 
\usepackage{verbatim}
\usepackage{multirow}
\usepackage{subfigure}

\usepackage[T1]{fontenc}
\usepackage[utf8]{inputenc}

\usepackage{footnote}

\makesavenoteenv{tabular}
\makesavenoteenv{table}
\usepackage{supertabular}
\usepackage{threeparttable}
\usepackage{longtable}
\usepackage{rotating}
\usepackage{lscape}

\usepackage{lineno,hyperref}
\modulolinenumbers[5]
\hypersetup{colorlinks,citecolor=blue,linkcolor=blue,urlcolor=blue}

\usepackage[]{algorithm2e}
\usepackage{dsfont}
\usepackage{siunitx}

\usepackage{float}

\usepackage{listings}
\usepackage{mdframed}

\lstnewenvironment{codebkg}{%
  \lstset{
    escapechar=\&,
    basicstyle=\ttfamily\scriptsize,
    backgroundcolor=\color{lightgray},
    frame=single,
    breaklines=true,    
    framerule=0pt,
    columns=fullflexible
  }
}
{}

\lstnewenvironment{texthighlight}{%
  \lstset{
    escapechar=\&,
    basicstyle=\ttfamily\scriptsize,
    backgroundcolor=\color{lightgray},
    frame=single,
    breaklines=true,  
    breakindent=0em,
    framerule=0pt,
    columns=fullflexible
  }
  \lstset{
    moredelim=[is][\itshape]{**}{**}
  }
}
{}

\newcommand{\red}[1]{\textcolor{red}{#1}}

\definecolor{babypink}{rgb}{0.96, 0.76, 0.76}
\definecolor{beaublue}{rgb}{0.74, 0.83, 0.9}
\definecolor{mediumturquoise}{rgb}{0.28, 0.82, 0.8}
\definecolor{mossgreen}{rgb}{0.68, 0.87, 0.68}
\definecolor{mustard}{rgb}{1.0, 0.86, 0.35}
\definecolor{olivine}{rgb}{0.6, 0.73, 0.45}
\definecolor{orchid}{rgb}{0.85, 0.44, 0.84}
\definecolor{palecerulean}{rgb}{0.61, 0.77, 0.89}
\definecolor{palegold}{rgb}{0.9, 0.75, 0.54}
\definecolor{paleplum}{rgb}{0.8, 0.6, 0.8}
\definecolor{palespringbud}{rgb}{0.93, 0.92, 0.74}
\definecolor{pastelgray}{rgb}{0.81, 0.81, 0.77}
\definecolor{pastelviolet}{rgb}{0.8, 0.6, 0.79}
\definecolor{pastelred}{rgb}{1.0, 0.41, 0.38}
\definecolor{pearl}{rgb}{0.94, 0.92, 0.84}
\definecolor{pistachio}{rgb}{0.58, 0.77, 0.45}	
\definecolor{teal}{rgb}{0.0, 0.5, 0.5}
\definecolor{tiffanyblue}{rgb}{0.04, 0.73, 0.71}
\definecolor{turquoise}{rgb}{0.19, 0.84, 0.78}
\definecolor{verdigris}{rgb}{0.26, 0.7, 0.68}
	
\definecolor{lightgray}{gray}{0.9}
\newcommand{\blue}[1]{\textcolor{blue}{#1}}

\newcommand{\hii}{H\textsc{ii}}

\title{\emph{radio-llava}: Advancing Vision-Language Models for Radio Astronomical Source Analysis}

\author{S. Riggi}
\affiliation{INAF - Osservatorio Astrofisico di Catania, Via Santa Sofia 78, 95123 Catania, Italy}
\email[S. Riggi]{simone.riggi@inaf.it}

\author{T. Cecconello}
\affiliation{INAF - Osservatorio Astrofisico di Catania, Via Santa Sofia 78, 95123 Catania, Italy}
\alsoaffiliation{Department of Electrical, Electronic and Computer Engineering, University of Catania, Santa Sofia 64, 95123, Catania, Italy}

\author{A. Pilzer}
\affiliation{NVIDIA AI Technology Center, Italy}

\author{S. Palazzo}
\affiliation{Department of Electrical, Electronic and Computer Engineering, University of Catania, Santa Sofia 64, 95123, Catania, Italy}

\author{N. Gupta}
\affiliation{CSIRO Space \& Astronomy, PO Box 1130, Bentley WA 6102, Australia}

\author{A.M. Hopkins}
\affiliation{School of Mathematical and Physical Sciences, 12 Wally’s Walk, Macquarie University, NSW 2109, Australia}

\author{C. Trigilio}
\affiliation{INAF - Osservatorio Astrofisico di Catania, Via Santa Sofia 78, 95123 Catania, Italy}

\author{G. Umana}
\affiliation{INAF - Osservatorio Astrofisico di Catania, Via Santa Sofia 78, 95123 Catania, Italy}

\begin{document}
\sloppy

\begin{abstract} 
The advent of next-generation radio telescopes is set to transform radio astronomy by producing massive data volumes that challenge traditional processing methods. Deep learning techniques have shown strong potential in automating radio analysis tasks, yet are often constrained by the limited availability of large annotated datasets. Recent progress in self-supervised learning has led to foundational radio vision models, but adapting them for new tasks typically requires coding expertise, limiting their accessibility to a broader astronomical community. Text-based AI interfaces offer a promising alternative by enabling task-specific queries and example-driven learning. In this context, Large Language Models (LLMs), with their remarkable zero-shot capabilities, are increasingly used in scientific domains. However, deploying large-scale models remains resource-intensive, and there is a growing demand for AI systems that can reason over both visual and textual data in astronomical analysis.

This study explores small-scale Vision-Language Models (VLMs) as AI assistants for radio astronomy, combining LLM capabilities with vision transformers. We fine-tuned the LLaVA VLM on a dataset of 59k radio images from multiple surveys, enriched with 38k image-caption pairs from the literature. The fine-tuned models show clear improvements over base models in radio-specific tasks, achieving $\sim$30\% F1-score gains in extended source detection, but they underperform vision-only classifiers and exhibit $\sim$20\% drop on general multimodal tasks. Inclusion of caption data and LoRA fine-tuning enhances instruction-following and helps recover $\sim$10\% accuracy on multimodal benchmarks (e.g., ChartQA/DocVQA).

This work lays the foundation for future advancements in radio VLMs, highlighting their potential and limitations, such as the need for better multimodal alignment, higher-quality datasets, and mitigation of catastrophic forgetting.
\end{abstract}


\section{Introduction}
\label{sec:intro}
The upcoming Square Kilometer Array (SKA) \citep{SKADesignDoc} and its precursor telescopes (e.g., MeerKAT, ASKAP, LOFAR) are revolutionizing radio astronomy, enabling to probe the radio sky at unprecedented sensitivities and angular resolutions. 
SKA, once operational, is expected to produce exabytes of data annually. The immense volume and complexity of the generated data will challenge traditional data-processing methods, necessitating advanced computational and AI techniques to automate repetitive, resource-intensive tasks. 

In this context, deep-learning methodologies have already shown promising results in various analysis tasks including: source detection in 2D radio maps \citep{Mostert2022,Zhang2022,Yu2022,RiggiMaskRCNN,Lao2023,Cornu2024,Stuardi2024}, source and host galaxy detection from 2D radio+IR maps \citep{Wu2019,Gupta2023} or radio+optical maps \citep{Lou2023}, source detection in HI cubes \citep{Liang2023,Hakansson2023,Barkai2023}, source classification \citep{Aniyan2017,Tang2019,Ma2019,Maslej2021,Tang2022,Nair2022,Riggi2024a}, search for objects with peculiar morphology \citep{Ralph2019,Galvin2020,Mostert2021,Gupta2022,Mesarcik2023,Lochner2023,Riggi2024b}, fast radio burst detection \citep{Connor2018,Agarwal2020}, radio imaging \citep{Schmidt2022,Geyer2023,Chiche2023}, synthetic data generation \citep{Rustige2023,Sortino2024,Martinez2024}, and many others. The full potential of developed models, especially those using supervised learning techniques, is often hampered by the scarcity of large and balanced annotated radio datasets. Additionally, existing radio models typically employ data labelling schemes that vary widely across different analysis cases, hindering the integration of individual datasets into larger collections and restricting model usability beyond their initial applications. 

Recent studies \citep{Slijepcevic2024,Riggi2024b,Lastufka2024} have sought to address the challenges posed by limited annotated training datasets by applying self-supervised learning (SSL) techniques, which utilize the extensive collections of unlabelled radio images available in current and past surveys.
Several foundational radio models have been developed to effectively enable feature extraction from radio maps for a variety of tasks, such as data inspection, source extraction and classification, anomaly detection, and image retrieval. Pretrained SSL models have also been fine-tuned on smaller annotated datasets and specialized for these applications. Ongoing research is focusing on various areas: comparing alternative SSL methods on radio data \citep{CecconelloPRRS2024}, assessing the performance of SSL models pretrained on non-radio data (such as natural or optical images) for radio-specific tasks and vice versa \citep{RiggiADASS2024,Lastufka2024b}, exploring optimized dataset curation strategies, scaling up model training to larger architectures and millions of radio images, and defining more constraining downstream datasets and tasks.

While existing SSL models can be adapted or expanded for new use cases, their accessibility is often limited by the need for astronomers to write code for adaptation to similar or entirely new tasks.
This requirement could hinder widespread adoption, as many astronomers may prefer more intuitive, user-friendly interfaces. An AI assistant with a more accessible, text-based interface would allow researchers to interact with the model by providing examples, querying specific tasks, and customizing output formats to suit their needs.

Large Language Models (LLMs) like GPT-4 \citep{GPT4}, Claude3 \citep{Claude3}, and open-source alternatives such as LLaMA \citep{LLaMA} or InternLM \citep{InternLM} have proven effective as AI assistants, showing remarkable zero-shot learning capabilities when prompted unseen data or tasks across a wide array of fields, including astronomy \citep{Tanoglidis2024}.  
Specializing and deploying very large open-source models is, however, currently prohibitive in terms of the required computing resources (high GPU requirements, memory demands, and power consumption). Furthermore, in addition to textual interaction, there is an increasing demand for models capable of processing visual data, facilitating multimodal reasoning for tasks like analysing complex astronomical images.
Some initiatives, such as AstroLLaMA \citep{Nguyen2023,Perkowski2024}, have started to address this by developing astronomy-specific mid-size LLMs, though these efforts are still limited to text-based inputs. Commercial solutions add additional cost concerns, particularly in inference and fine-tuning expenses, as demonstrated by \cite{Sun2024} in the context of interpreting multi-band galaxy observations. Furthermore, while large models are well-suited for tasks requiring extensive general knowledge, this scope may be more than what is needed in astronomy, where specialized knowledge is essential. Given these considerations, current research has increasingly turned to adapting smaller LLMs (i.e., those with fewer than 10 billion parameters) for specific domains, as well as investigating multi-modal models capable of processing combined data inputs, such as text, images, and videos.

In this context, we aim to explore recent, state-of-the-art, small-scale vision-language models (VLMs) to develop AI assistants tailored to radio astronomy. These models combine both visual and textual comprehension by integrating LLM capabilities with vision transformers \citep{Dosovitskiy2021} for image processing. 
Typically, VLMs comprise two main components: a vision encoder transformer that extracts features from input images, and an LLM that generates textual responses from combined visual and textual input representations.
VLMs offer promising solutions to the outlined challenges in two ways. First, they can manage tasks through a text-based interface, allowing astronomers to specify task details and expected response formats in natural language. This flexibility enables the use of contextual image-text examples, making VLMs more adaptable than traditional vision models that are limited to predefined label outputs for specific tasks. Second, VLMs potentially support the integration of specialized vision encoder models that have been trained on unlabelled radio data through self-supervised learning (SSL) methods. This approach bridges the gap with ongoing SSL research, facilitating the reuse of existing radio SSL models and enabling the full utilization of extensive unlabelled image datasets from recent and past radio surveys.
These smaller models offer a promising alternative, with lower computational costs and more manageable deployment requirements. Moreover, their specialized capabilities could be more than sufficient for the specific needs of radio astronomy, without the added complexity and resource demands of larger models. However, the suitability of these smaller models for astronomical tasks remains largely unexplored, particularly in radio source analysis.

This paper seeks to evaluate the current state of small multimodal language models as AI assistants for radio astronomy. By investigating their strengths, limitations, and applicability to radio astronomical source analysis tasks, we hope to familiarize the community with this emerging technology and its potential, as well as to highlight the challenges that need to be addressed in future developments.
A tailored VLM for radio astronomy could eventually assist astronomers in efficiently analysing radio images without requiring extensive technical expertise in AI models. By leveraging a text-based interface, astronomers can perform complex and diverse image analysis tasks, even guiding the model using image-based examples. Key applications include the automated identification and retrieval of specific classes of radio sources in survey image data, as well as data quality assessment - enhancing the efficiency and accessibility of radio survey analysis. 
Additionally, a VLM-based assistant could be deployed to support less experienced users (e.g., students, citizens) in ongoing crowdsourcing projects like the EMU Radio Galaxy Zoo\footnote{\url{https://www.zooniverse.org/projects/hongming-tang/radio-galaxy-zoo-emu}}. By providing real-time guidance and explanations, the model could help users classify radio sources, identify peculiar objects, and improve the reliability of crowdsourced annotations.

Multi-modal models have only very recently begun to gain traction in astronomy - with most developments emerging within the past year - as a means to bridge heterogeneous data modalities such as images, spectra, and natural language. These efforts have primarily focused on adapting the CLIP (Contrastive Language–Image Pretraining) model \citep{Radford2021} and its derivatives to astronomical tasks, leveraging their ability to align visual and textual representations in a shared latent space. For instance, \cite{Gupta2025} introduced \texttt{EMUSE}, a tool built on a fine-tuned OpenCLIP\footnote{\url{https://github.com/mlfoundations/open_clip}} model that enables users to search EMU survey data using either textual queries or template image similarity. A similar application is \texttt{PAPERCLIP} \citep{PAPERCLIP}, a CLIP-based model fine-tuned on Hubble Space Telescope (HST) proposal data $-$ including image observations and proposal abstracts $-$ which enables cross-modal retrieval based on image content or textual queries such as object names or scientific use cases.
AstroCLIP \citep{AstroCLIP} presents a powerful CLIP-style foundation model for galaxies in the optical domain, where the textual modality in the original CLIP framework is replaced by galaxy spectra. The model aligns image and spectrum modalities in a shared embedding space after self-supervised pre-training of each modality independently, and demonstrates impressive performance on tasks such as morphology classification, redshift estimation, and property inference.

Unlike these studies, our approach focuses on instruction-following vision-language models like LLaVA (\emph{Large Language and Vision Assistant}) \citep{LLaVA} to enable open-ended captioning, visual question answering, and multi-turn scientific dialogue grounded in domain-specific radio astronomy data. These tasks go beyond the static alignment capabilities of CLIP models, which lack generative, reasoning, and conversational abilities. Our model is thus particularly suited for exploratory analysis, educational interfaces, and assistant-style tools that can explain, summarise, or discuss diagnostic plots and observational data. Meanwhile, CLIP-based models remain better suited for scalable retrieval, zero-shot classification, and semantic similarity search over large datasets.

The paper is organized as follows. Section~\ref{sec:vlm} provides an overview of vision-language models, with a focus on the architecture of a prominent model, LLaVA, which we aim to adapt for radio-astronomical data. In Section~\ref{sec:radio-llava}, we describe our adapted model, termed \emph{radio-llava}, including the training datasets and methodology. 
Section~\ref{sec:model-evaluation} presents the evaluation of the specialized model across several radio-astronomy tasks. Finally, Section~\ref{sec:summary} summarises the results and discusses directions for future research.

\section{Vision-Language Models}
\label{sec:vlm}
Multi-modal large language models (MLLMs) are designed to process and integrate data from multiple modalities, such as text, audio, images, and video. Vision-language models (VLMs) are a specific type of multi-modal system that focuses on combining visual and textual information. These models leverage large language models and vision transformers to align visual and textual representations, enabling them to perform complex tasks like image captioning, visual question answering (VQA), and object recognition in a descriptive context. A comprehensive review of MLLMs and VLMs is provided by \cite{Li2023,Bordes2024,Yin2024}. In this section, we focus on describing the current state, architecture, and training strategy of the VLM model \emph{LLaVA}, which we have adapted for use with radio astronomical data in this work.

\subsection{The LLaVA model}

\subsubsection{Model overview}
LLaVA (\emph{Large Language and Vision Assistant}) \citep{LLaVA} is a state-of-the-art multimodal model that integrates both visual and textual understanding, combining the capabilities of large language models (LLMs) with vision processing abilities. Its primary function is to interpret and generate responses to input that includes both images and text, making it ideal for tasks like visual question answering (VQA), image captioning, and other vision-language tasks. Since the first release, the model demonstrated exceptional multimodal conversational skills, often displaying behaviour comparable to GPT-4V when tasked with interpreting novel images and following new instructions for the first time.

Following releases (LLaVA 1.5, \citealt{LLaVA_1.5}) greatly enhanced model capabilities by integrating a larger set of academic-focused instructional data, achieving state-of-the-art results on numerous benchmarks while utilizing a highly data-efficient strategy. Recent advancements in the LLaVA series, including models like LLaVA-NeXT \citep{LLaVA-NeXT} and LLaVA-OneVision \citep{LLaVA-OneVision}, have significantly broadened the scope of input modalities they can handle, supporting both single or multiple images as well as video content. These improvements were driven by three key innovations: the \emph{AnyRes} technique for processing high-resolution images, the expansion of high-quality instruction datasets, and the integration of the most advanced open-source LLMs available at the time, further enhancing model capabilities across diverse tasks. Various variants or specialization of the first LLaVA models have been produced so far. For example, TinyLLaVA \citep{TinyLLaVA,TinyLLaVA-Factory} is a compact refactored variant of the original LLaVA 1.5 model, designed to enable easier inclusion of alternative light vision and LLM models, thus significantly reducing overall model size and resource requirements. LLaVA-Med \citep{LLaVA-Med} is a specialized variant of the LLaVA model designed to assist in medical image analysis and diagnostics by fine-tuning its multimodal capabilities on medical datasets such as X-rays, MRIs, and other healthcare-related visual data.

\begin{figure}[htb]
\centering%
\includegraphics[scale=0.7]{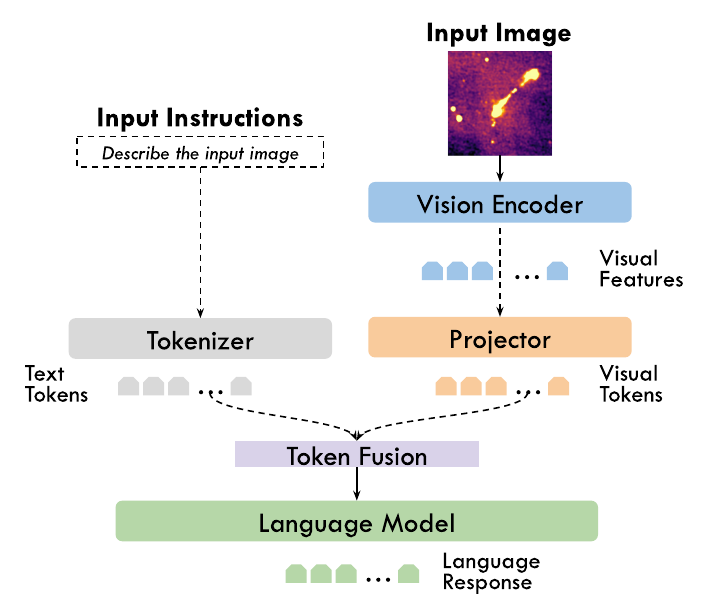}
\vspace{-0.2cm}
\caption{A schematic representation of the LLaVA model architecture.}%
\label{fig:llava-architecture}
\end{figure}

\subsubsection{Model architecture}
The LLaVA model, schematically represented in Figure~\ref{fig:llava-architecture}, consists of these components:
\begin{itemize}
\item \emph{Vision Encoder}: Processes image data using a pre-trained Vision Transformer (ViT) model with multiple transformer layers, such as CLIP \citep{Radford2021} or SigLIP \citep{Zhai2023b}. It extracts visual features from input images;
\item \emph{Language Model}: Handles text processing, typically an LLM such as Qwen \citep{Bai2023}, capable of understanding, generating, and reasoning with natural language;
\item \emph{Projector}: Since the vision encoder and language model operate in different feature spaces, the projector transforms visual embeddings into a format compatible with the language model input space. This enables effective integration of visual and textual information. LLaVA employs a two-layer Multi-Layer Perceptron (MLP) trained to align the modalities, ensuring that the visual embeddings can be seamlessly used by the language model;
\item \emph{Multimodal Fusion Layer}: 
It aligns and fuses the visual features from the vision encoder with the text embeddings from the language model, enabling the model to process both modalities jointly through self-attention mechanisms.
\end{itemize}
The model processes multimodal image-text inputs as follows:
\begin{enumerate}
\item The input text (instruction or query) is tokenized into numerical tokens using a predefined vocabulary.
\item The input image is divided into patches and passed through the vision encoder, which extracts key features such as objects, colours, textures, and spatial relationships. The projector then converts these visual representations into language-compatible embeddings;
\item The multimodal fusion layer integrates the visual embeddings into the input sequence of the language model, allowing it to process both visual and textual data jointly. The model then generates an output response based on the given task, such as answering questions about the image (VQA) or generating a descriptive caption (image captioning).
\end{enumerate}

\subsubsection{Model training}
The LLaVA model series is trained from pre-existing language and vision encoders through instruction fine-tuning on large-scale datasets. These datasets consist of image (or video) and text pairs, including captions, descriptions, and questions, enabling the model to learn associations between visual elements and natural language. The training process typically involves multiple stages, each potentially using different datasets, including pretraining on unimodal visual and textual data, aligning vision and language features, and fine-tuning with instructional data to address diverse visual tasks. Further details on training datasets and methodology can be found in the original model publications.

During instruction tuning, the model is optimized by minimizing a cross-entropy loss, which quantifies how closely the predicted text output matches the ground truth. The model generates text auto-regressively, predicting one token at a time based on previously generated tokens. At each step, it outputs a probability distribution over possible next tokens and is trained to minimize the difference between its prediction and the actual token. The cross-entropy loss is computed for each token while conditioning on prior tokens, accumulating over the entire sequence and penalizing incorrect predictions at each step. This iterative process ensures the model learns to generate coherent text in a structured manner.

\section{The \emph{radio-llava} model}
\label{sec:radio-llava}
The \emph{radio-llava} model is a small multi-modal model fine-tuned from a base LLaVA model using radio astronomical image-text data. This section describes the training dataset, and the model fine-tuning procedure.

\subsection{Training dataset}
\label{subsec:training-datasets}
The training dataset consists of multiple conversations between a virtual assistant and a user regarding a given radio image. The dataset follows the standard JSON format required by multi-modal models:
\begin{codebkg}
[
  {
    "id": "image id",
    "image": "image path",
    "conversations": [
      {
        "from": "human",
        "value": "<image>\n Provide a brief description of the given image."
      },
      {
        "from": "gpt",
        "value": "The image is a radio astronomical cutout ..."
      },
      ...
]
\end{codebkg}
We constructed two training datasets. The first, referred to as the \emph{Q\&A dataset}, consists of a series of question-answer interactions related to the content of radio images, all extracted from radio continuum surveys. The second dataset, termed the \emph{caption dataset}, contains a single user-assistant exchange per image, in which the assistant provides a description of the image content. In this case, images and their corresponding captions were sourced from a collection of scientific papers on radio astronomical topics available in the arXiv database. Details on both datasets are provided in the following sections.

\subsubsection{Q\&A dataset}
\label{subsec:training-datasets-qa}
This dataset was assembled from multiple annotated radio datasets, each designed for different radio source classification or detection tasks:
\begin{itemize}
\item \textit{Fine-Grained Datasets}: These datasets, typically used for training radio object detection and segmentation models like YOLO \citep{YOLO} or Mask R-CNN \citep{MaskRCNN}, contain wide-field images (a few arcminutes in size) with region- or pixel-level annotations, including object positions (centres, bounding boxes, segmentation masks), classification labels, and confidence scores.
\item \textit{Coarse-Grained Datasets}: Commonly used for radio source classification models, these datasets contain either zoomed-in source images or wide-field images with one or more assigned classification labels.
\end{itemize}
Details regarding the number of images and available classes for each dataset are provided in Appendix~\ref{appendix:training-datasets}.

It is important to note that classification schemes vary across datasets. Some provide astrophysical source-type labels (e.g., \texttt{HII}, \texttt{SNR}, \texttt{GALAXY}), while others focus on morphological classifications (e.g., \texttt{FR-I} vs. \texttt{FR-II}, \texttt{COMPACT} vs. \texttt{EXTENDED} vs. \texttt{DIFFUSE}). Before generating the conversational Q\&A dataset, we aimed to standardize terminology whenever possible. In some cases (e.g., see Section~\ref{subsec:fine-grained-datasets}), we performed cross-matching and relabelling to augment the original datasets by adding additional labels to certain images or objects. However, variations in labelling schemes remain unavoidable due to the lack of annotation standards in the radio astronomy community. In this respect, our goal is to fine-tune an LLM model that is exposed to diverse classification schemes, making it potentially adaptable to different user domains.

The Q\&A dataset was constructed from annotated datasets through the following steps:
\begin{enumerate}
\item \textit{Automated Template-Based Descriptions}: For each image, we programmatically generated a template description based on the available annotations\footnote{These descriptions are statically defined, meaning two images with identical annotations will have the same description.}.

Example: An image from the \textit{radioimg-dataset} (see Section~\ref{subsec:radioimg-multilabel-dataset}) with the assigned labels {\texttt{COMPACT}, \texttt{EXTENDED}, \texttt{RADIO-GALAXY}, \texttt{ARTIFACT}} would be described as:
\begin{texthighlight}
**The image is a radio astronomical image cutout extracted from a larger radio-continuum Stokes-I map produced by an interferometer telescope. The image contains various point-like or compact radio sources superimposed over the sky background noise. It also contains one or more extended radio sources. Some of them are likely extended radio galaxies. Some radio sources present in the image are poorly imaged and surrounded by imaging artefacts having a ring pattern.**
\end{texthighlight}
Fine-grained datasets include richer descriptions, specifying source positions and sizes.
\item \textit{Automated Q\&A Generation Using a Pretrained VLM}: We generated multiple Q\&A interactions per image using a \texttt{InternVL} VLM model\footnote{We used \texttt{InternVL2\_5-8B-MPO} model version, available here: \url{https://huggingface.co/OpenGVLab/InternVL2_5-8B-MPO}, \url{https://github.com/OpenGVLab/InternVL}} \citep{InternVL}.
The model was fed with the image, template caption, and structured prompts to ensure that the generated conversations remained faithful to the original image and annotation content, and included at least the following questions:
\begin{itemize}
\item \textit{Can you describe the image content?}
\item \textit{Can you provide the bounding box coordinates of all radio sources with class $X$ (e.g., compact, extended, etc.) present in the image}?
\item \textit{Do you see any likely radio galaxy with an extended morphology in the image?}
\item \textit{Which of these morphological classes of radio sources do you see in the image?}
\item \textit{Do you see any imaging artefact around bright sources in the presented image?}
\item \textit{Is there any blank pixel region at the edges of the image?}
\item \textit{Is the image content ordinary or peculiar in terms of the objects it contains?}
\end{itemize}
To prevent excessive generalization, we constrained the VLM's output by using an intermediate temperature setting (0.5).
\end{enumerate}
Overall, the final training dataset comprises 59,331 images and 1,590,202 user-assistant conversations.

Despite these efforts, the current annotated radio data $-$ which primarily provide classification labels or bounding boxes $-$ still limit the diversity and richness of generated image-based conversations. This constraint directly impacts model performance and its instruction-following capabilities, as discussed in Section~\ref{sec:model-evaluation}.

\subsubsection{Caption dataset}
\label{subsec:training-datasets-caption}
This dataset was compiled by extracting figures and their corresponding captions from a broad collection of arXiv scientific papers containing radio astronomy-related keywords, published between 2000 and 2025. To classify the image format and assess caption quality, we processed the extracted raw image-caption data using the same \texttt{InternVL} VLM model employed for generating the Q\&A dataset. Specifically, we computed the following parameters for each image-caption pair:
\begin{itemize} 
\item \texttt{n\_words}: number of words in the caption; 
\item \texttt{has\_multiplot}: binary flag set to \texttt{true} if the image contains multiple plots/frames, either as insets, side by side, stacked, or arranged in a grid layout; 
\item \texttt{is\_astromap}: binary flag set to \texttt{true} if the image and caption depict an astrophysical map with one or more sources superimposed on the sky background; 
\item \texttt{is\_corrupted}: binary flag set to \texttt{true} if the caption contains incomplete sentences or corrupted text; 
\item \texttt{caption\_score}: integer score assessing caption quality on a scale from 0 (low) to 10 (high), based on coherence, informativeness, completeness, clarity, and correctness of English style. 
\end{itemize}
Only highly rated single-plot images were included in the training sample, applying the following selection criteria: \texttt{n\_words}>5, \texttt{has\_multiplot}=0, \texttt{is\_corrupted}=0, and \texttt{caption\_score}>7. This resulted in a final training sample of 38,545 images. The \texttt{has\_multiplot} criterion had a significant impact, removing approximately 62\% of the images from the initial dataset. We opted not to apply the \texttt{is\_astromap} filter, as doing so would have further reduced the dataset size to approximately 8,700 images.

\subsection{Model fine-tuning}
The \emph{radio-llava} model was trained using instruction fine-tuning on the Q\&A radio dataset alone, as well as on the combined Q\&A and caption datasets, starting from the pre-trained LLaVA-OneVision 7B model\footnote{\url{lmms-lab/llava-onevision-qwen2-7b-ov}}. Keeping the vision encoder (\texttt{siglip-so400m-patch14-384}) frozen, we fine-tuned both the LLM (\texttt{qwen2}) and adapter (\texttt{mlp2x\_gelu}) components using either full fine-tuning or the Low-Rank Adaptation (LoRA) \citep{LORA} fine-tuning strategy\footnote{LoRA is a lightweight training method that updates only small, low-rank matrices within the model instead of fine-tuning the entire model. This significantly reduces computational overhead and storage requirements while maintaining high performance}. We set the LoRA rank and alpha scaling parameters to 64 and 128, respectively. The model was trained for either 1 epoch (\textit{shallow} fine-tuning) or 3 epochs (\textit{deep} fine-tuning). In all training runs, we used default hyperparameters, with a batch size of 1, a gradient accumulation step of 2, and a learning rate of 10$^{-5}$.

On single-GPU servers with medium GPU memory (e.g., NVIDIA A30 24 GB or RTX6000 48 GB), we were only able to train the model using LoRA fine-tuning on the Q\&A dataset, while full fine-tuning required more extensive computational resources. Consequently, all fine-tuning runs were conducted on a single node of the CINECA LEONARDO Booster infrastructure\footnote{\url{https://www.hpc.cineca.it/systems/hardware/leonardo/}}, utilizing 4 GPUs (NVIDIA A100 64 GB) and 8 CPUs (Intel Xeon Platinum 8358, 2.60GHz) with 32 GB of memory allocated per CPU. Full fine-tuning required approximately 15 hours per epoch on the combined training dataset.

\subsection{Implementation details}
For training the LLaVA model, we utilized software and resources available in the LLaVA-NeXT repository\footnote{\url{https://github.com/LLaVA-VL/LLaVA-NeXT}}. Only minor modifications were applied to the original software to load our dataset and enable the loading of LoRA fine-tuned models using the Qwen LLM.

Inference scripts developed for LLaVA and other VLM models tested in this study are provided in the following repository: \url{https://github.com/SKA-INAF/radio-llava}. This repository also includes a Streamlit\footnote{\url{https://streamlit.io/}} application (see Figure~\ref{fig:streamlit-app} in the Appendix), allowing users to load a LLaVA model, upload an image, and interact with the assistant via a web interface. Table~\ref{tab:llava-conversation-example} in the Appendix provides an example of user-assistant conversations for two sample radio images, comparing responses from the base and fine-tuned LLaVA-OneVision 7B models.

The fine-tuned models are publicly available in the \emph{Hugging Face} repository: \url{https://huggingface.co/inaf-oact-ai}.

\section{Model evaluation}
\label{sec:model-evaluation}
Using independently annotated datasets, we defined six evaluation benchmarks (B1$-$B6) to assess the model's reasoning capabilities on radio image data. The benchmark datasets and inference prompts are detailed in Section~\ref{subsec:evaluation-benchmarks-radio}. Additionally, we evaluated our models on various standard non-astronomical benchmarks, listed in Section~\ref{subsec:evaluation-benchmarks-standard}, to quantify the impact of fine-tuning on tasks previously learned by the base model.

Section~\ref{subsec:analysis-zeroshot} presents the zero-shot performance of the LLaVA base model, comparing it with alternative VLMs. The evaluation results for the fine-tuned \emph{radio-llava} models are reported in Section~\ref{subsec:analysis-finetuning}.

We will consistently use these widely adopted metrics in classification problems: 
\begin{itemize} 
\item Recall ($\mathcal{R}$): The fraction of sources (images) from a given class that are correctly identified by the model, out of all sources (images) that truly belong to that class: 
\begin{equation*} 
\mathcal{R} = \frac{TP}{TP + FN} 
\end{equation*}
\item Precision ($\mathcal{P}$): The fraction of sources (images) correctly predicted to belong to a given class, out of all sources (images) the model assigned to that class: 
\begin{equation*} 
\mathcal{P} = \frac{TP}{TP + FP} 
\end{equation*}
\item F1-score: The harmonic mean of precision and recall, offering a balanced measure of a model's performance: 
\begin{equation} 
\text{F1-score} = 2 \times \frac{\mathcal{P} \times \mathcal{R}}{\mathcal{P} + \mathcal{R}} \end{equation}
\item Accuracy ($\mathcal{A}$): The overall fraction of correctly classified sources (images), regardless of class, over the total number of instances: 
\begin{equation*} 
\mathcal{A} = \frac{TP + TN}{TP + TN + FP + FN} 
\end{equation*} 
\end{itemize}
Here, $TP$, $FP$, $TN$, and $FN$ represent the number of true positives, false positives, true negatives, and false negatives, respectively.

\subsection{Evaluation benchmarks}
\label{subsec:evaluation-benchmarks}

\subsubsection{Radio benchmarks}
\label{subsec:evaluation-benchmarks-radio}

\paragraph{B1 - Extended/Diffuse Source Detection}
\label{subsec:smorph-evaluation}
We used the \texttt{radioimg-multilabel} test dataset (5,718 images) to evaluate the models' ability to detect extended or diffuse radio sources in input images. 

For this task, we applied the following prompt:
\begin{texthighlight}
### Context: Consider these morphological classes of radio astronomical sources, defined as follows: 
EXTENDED: This class comprises either single-island compact objects with sharp edges, having a morphology and size dissimilar to that of the image synthesised beam (e.g. 10 times larger than the beam size or with elongated shape), or disjoint multi-island objects, where each island can have either a compact or extended morphology and can host single or multiple emission components. Typical examples are extended radio galaxies formed by a single elongated island or by multiple islands, hosting the galaxy core and lobe structures
DIFFUSE: a particular class of single-island extended objects with small angular size (e.g. smaller than few arcminutes), having diffuse edges and a roundish morphology; 
DIFFUSE-LARGE: large-scale (e.g. larger than few arcminutes and covering a large portion of the image) diffuse object with irregular shape.
An island is a group of 4-connected pixels in an image under analysis with intensity above a detection threshold with respect to the sky background level.

### Question: Which of these morphological classes of radio sources do you see in the image?
EXTENDED
DIFFUSE
DIFFUSE-LARGE
Answer the question using the provided context (and examples). Report the identified class labels separated by commas, without any additional explanation text. Report just NONE if you cannot recognise any of the above classes in the image.
\end{texthighlight}

\paragraph{B2 - Source Morphology Classification}
\label{subsec:rgz-evaluation}
We considered the \texttt{rgz-smorph} test dataset, containing $\sim$3,835 images from the VLA FIRST survey,  each zoomed and centred around radio sources belonging to six distinct morphological classes: \texttt{1C-1P}, \texttt{1C-2P}, \texttt{1C-3P}, \texttt{2C-2P}, \texttt{2C-3P}, \texttt{3C-3P}. 

The model was evaluated using a single-label multi-class classification task with the following prompt:
\begin{texthighlight}
### Context: Consider these morphological classes of radio astronomical sources:
1C-1P: single-island sources having only one flux intensity peak;
1C-2C: single-island sources having two flux intensity peaks; 
1C-3P: single-island sources having three flux intensity peaks;
2C-2P: sources consisting of two separated islands, each hosting a single flux intensity peak;
2C-3P: sources consisting of two separated islands, one containing a single peak of flux intensity and the other exhibiting two distinct intensity peaks;
3C-3P: sources consisting of three separated islands, each hosting a single flux intensity peak. 
An island is a group of 4-connected pixels in an image under analysis with intensity above a detection threshold with respect to the sky background level.

### Question: Which of these morphological classes of radio sources do you see in the image?
1C-1P
1C-2C
1C-3P
2C-2P
2C-3P
3C-3P
Answer the question using the provided context (and examples). Report only the identified class label, without any additional explanation text.
\end{texthighlight}

\paragraph{B3 - Extended Radio Galaxy Detection}
\label{subsec:galaxydet-evaluation}
We used the \texttt{radioimg-multilabel} test dataset (5,718 images) to assess the models' ability to identify radio sources with morphologies characteristic of extended radio galaxies. 

For this task, we applied the following prompt:
\begin{texthighlight}
Do you see any likely radio galaxy with an extended morphology in the image?
Answer concisely: Yes or No.
\end{texthighlight}

\begin{figure*}[!htb]
\centering%
\subtable[B1]{\includegraphics[scale=0.45]{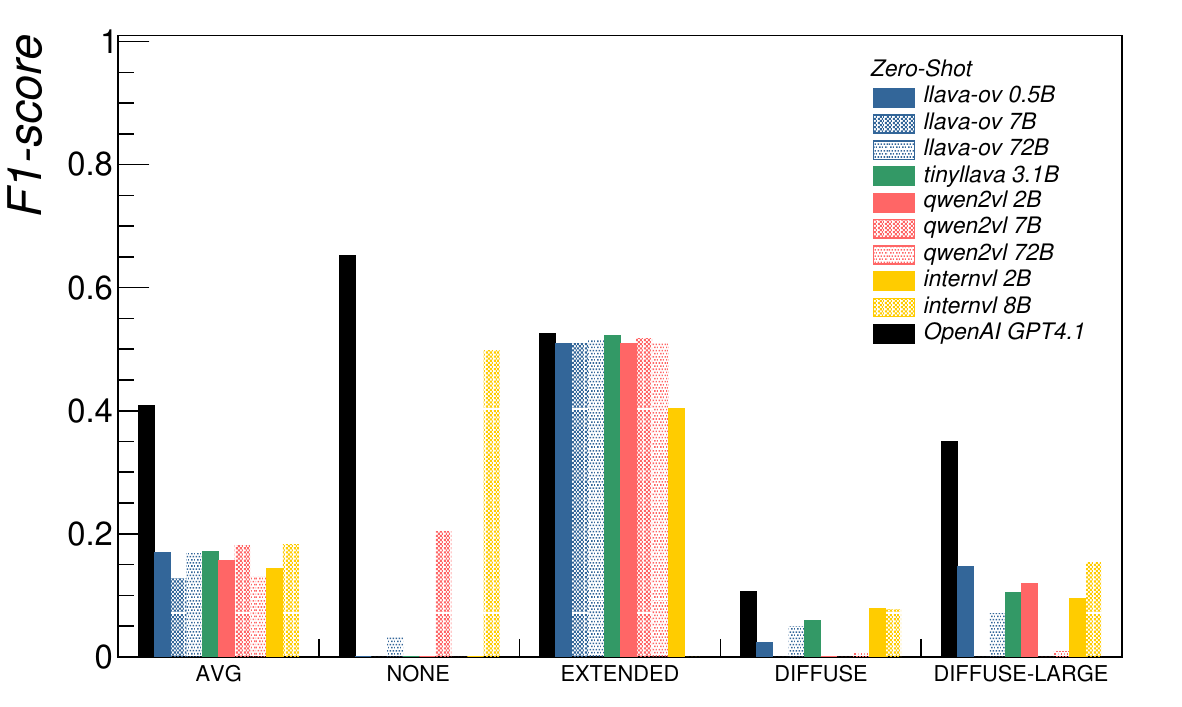}}
\hspace{-0.2cm}%
\subtable[B2]{\includegraphics[scale=0.45]{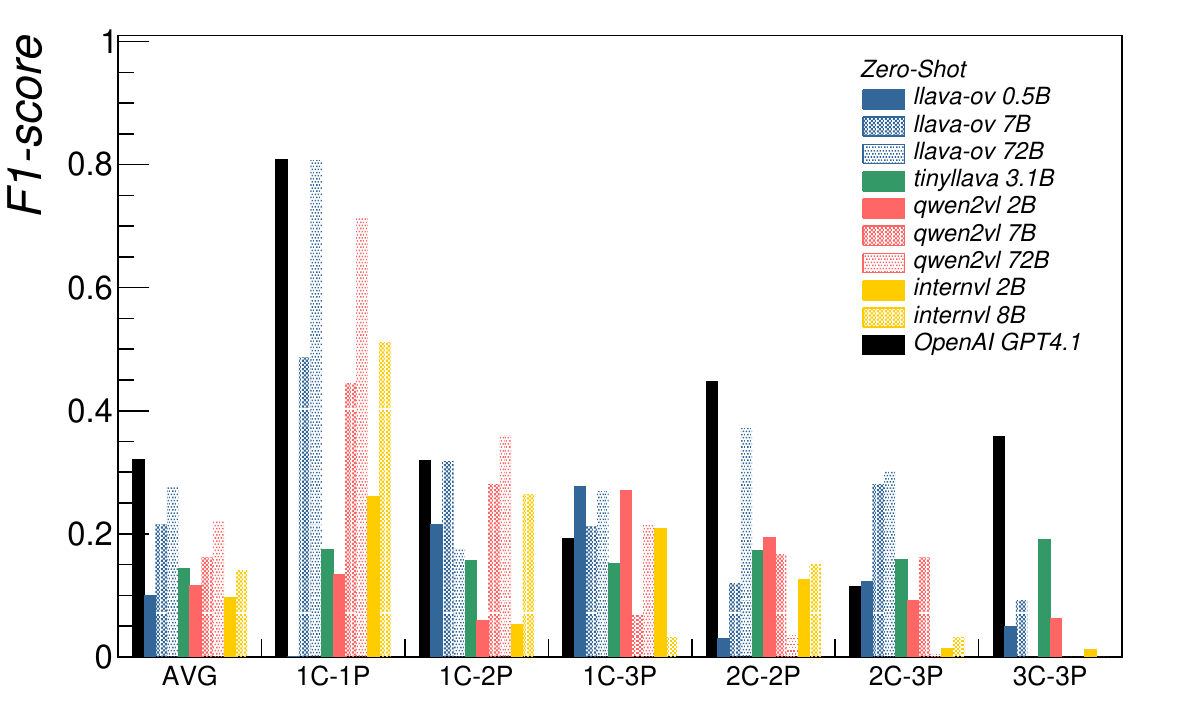}}\\
\vspace{-0.4cm}%
\subtable[B3]{\includegraphics[scale=0.45]{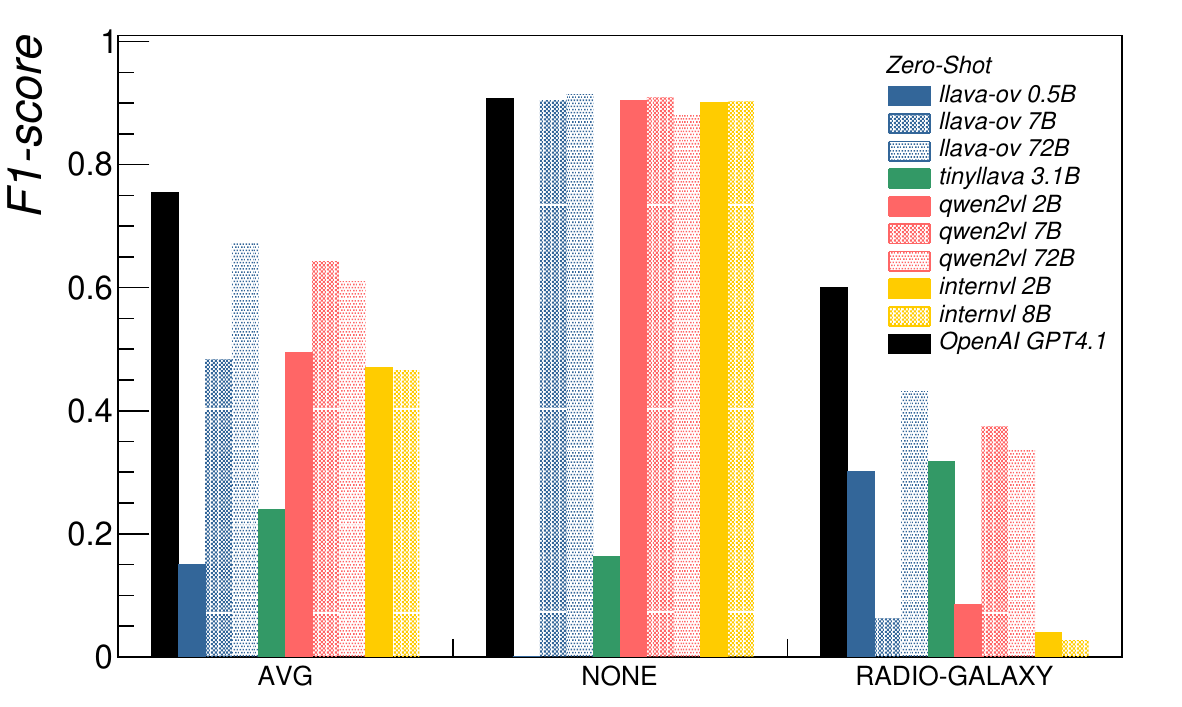}}
\hspace{-0.2cm}%
\subtable[B4]{\includegraphics[scale=0.45]{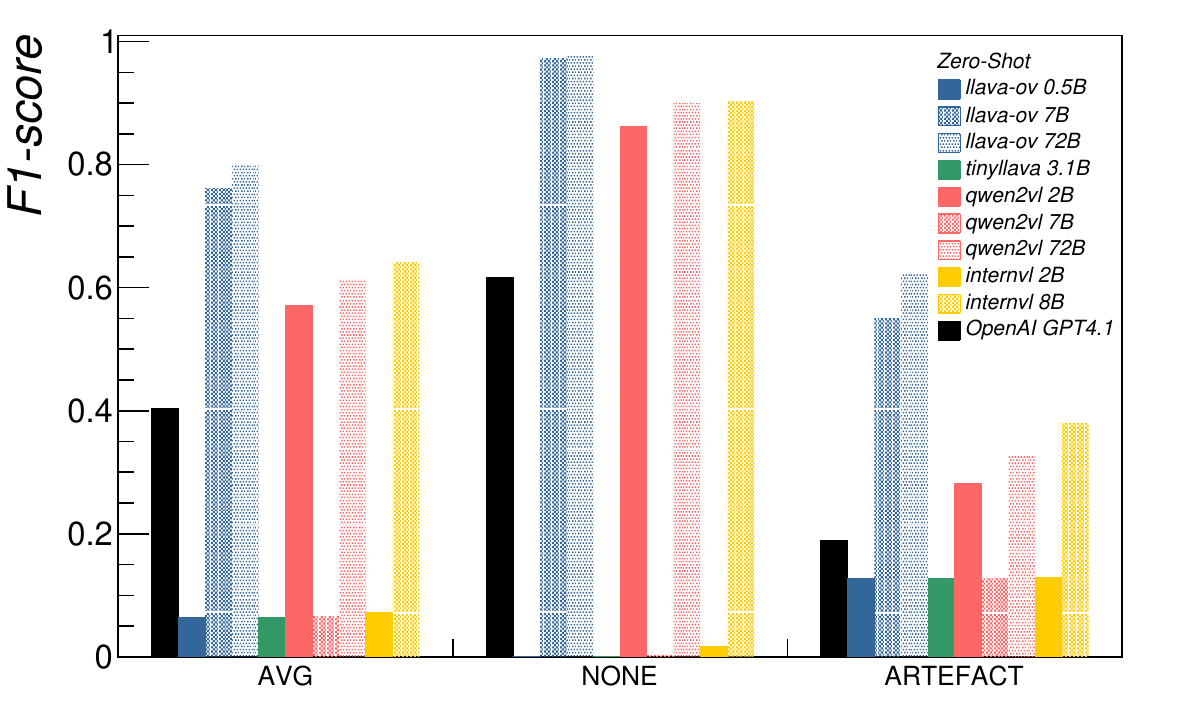}}\\
\vspace{-0.4cm}%
\subtable[B5]{\includegraphics[scale=0.45]{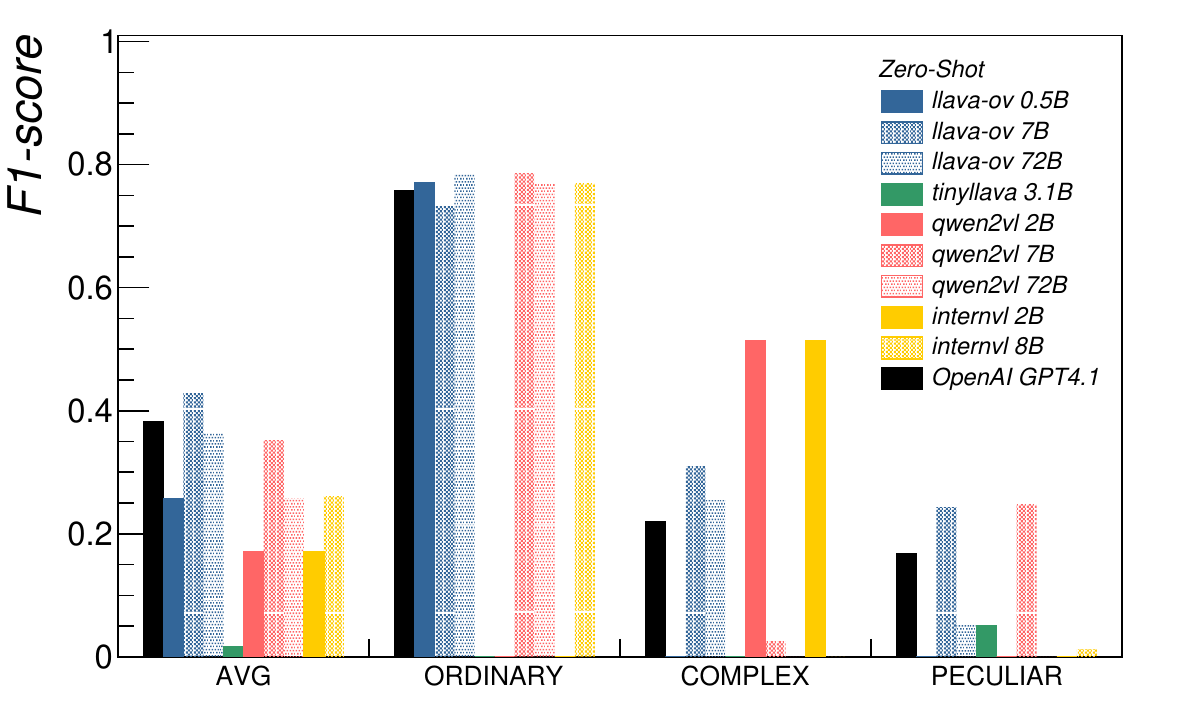}}
\hspace{-0.2cm}%
\subtable[B6]{\includegraphics[scale=0.45]{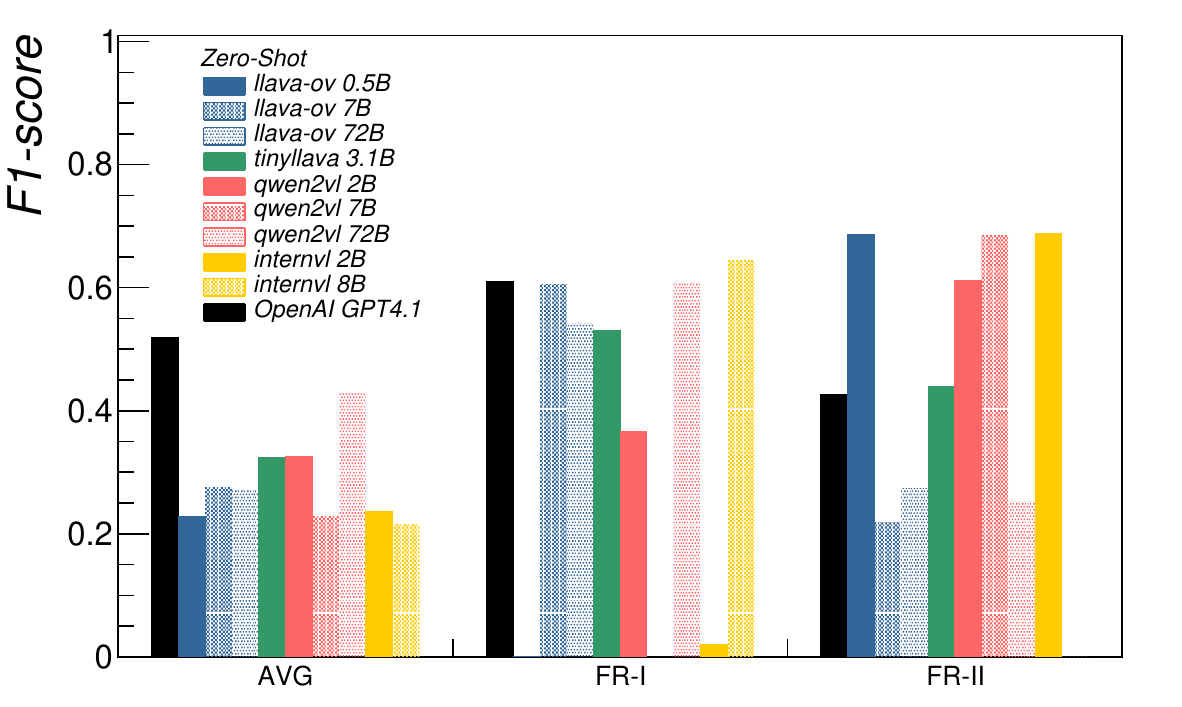}}
\vspace{-0.4cm}%
\caption{Classification F1-scores obtained with VLMs of different sizes (0.5B, 2B, 3.1B, 7B, 8B, 72B) in zero-shot mode over B1-B6 evaluation benchmarks. We report the F1-score for individual classes, as well as the class-averaged F1-score (labelled as 'AVG'). LLaVA, TinyLLaVA, Qwen2VL, and InternVL models are respectively shown with blue, green, red, and orange histograms. OpenAI GPT4.1 model is shown with black histograms.}%
\label{fig:eval-zeroshot}
\end{figure*}

\paragraph{B4 - Imaging Artefact Detection}
\label{subsec:artefactdet-evaluation}
We used the \texttt{radioimg-multilabel} test dataset (5,718 images) to evaluate the models' ability to detect imaging artefacts in input images. 

For this task, we considered the following prompt:
\begin{texthighlight}
Do you see any imaging artefact with a ring pattern around bright sources in the image?
Answer concisely: Yes or No.
\end{texthighlight}

\paragraph{B5 - Source Peculiar Morphology Classification}
\label{subsec:anomalydet-evaluation}
We used the \texttt{radioimg-multilabel} test dataset (5,718 images) to evaluate the models' ability to detect radio sources with complex or anomalous morphologies in input images. These sources were previously classified into three categories: \texttt{PECULIAR} (150 images), \texttt{COMPLEX} (1,978 images), and \texttt{ORDINARY} (3,590 images).

For this task, we applied the following prompt:
\begin{texthighlight}
### Context: Consider these radio image peculiarity classes, defined as follows: 
ORDINARY: image containing only point-like or slightly-resolved compact radio sources superimposed over the sky background or imaging artefact patterns;
COMPLEX: image containing one or more radio sources with extended or diffuse morphology;
PECULIAR: image containing one or more radio sources with anomalous or peculiar extended morphology, often having diffuse edges, complex irregular shapes, covering a large portion of the image.

### Question: Can you identify which peculiarity class the presented image belongs to? 
ORDINARY
COMPLEX
PECULIAR
Answer the question using the provided context (and examples). Report only the identified class label, without any additional explanation text.
\end{texthighlight}

\paragraph{B6 - Radio Galaxy Morphology Classification}
\label{subsec:galaxyclass-evaluation}
We used the \texttt{Mirabest} \citep{Mirabest} \emph{confident sample} dataset, which contains 833 images from the VLA FIRST survey, each zoomed and centred around radio galaxies belonging to two distinct morphological classes: \texttt{FR-I} (397 images) and \texttt{FR-II} (436 images).

For this task, we applied the following prompt:
\begin{texthighlight}
### Context: Consider these morphological classes of radio galaxies: 
FR-I: radio-loud galaxies characterised by a jet-dominated structure where the radio emissions are strongest close to the galaxy's centre and diminish with distance from the core;
FR-II: radio-loud galaxies characterised by an edge-brightened radio structure, where the radio emissions are more prominent in lobes located far from the galaxy's core, with hotspots at the ends of powerful, well-collimated jets.

### Question: Which of these morphological classes of radio galaxy do you see in the image?
FR-I
FR-II
Answer the question using the provided context (and examples). Report only the identified class label, without any additional explanation text.
\end{texthighlight}

\subsubsection{Image standard benchmarks}
\label{subsec:evaluation-benchmarks-standard}
We evaluated all \textit{radio-llava} fine-tuned models on 11 image benchmarks (\texttt{AI2}, \texttt{ChartQA}, \texttt{DocVQA}, \texttt{InfoVQA}, \texttt{MME}, \texttt{MMMU}, \texttt{MMStar}, \texttt{OCRBench}, \texttt{SEED-Bench}, \texttt{ScienceQA-IMG}, \texttt{RealWorldQA}), which are widely used to assess multimodal model performance across various tasks, ranging from diagram, chart, and scene understanding to text extraction. Further details on each benchmark are provided in Appendix~\ref{appendix:image-standard-benchmarks}.

\subsection{Zero-shot performance}
\label{subsec:analysis-zeroshot}
We evaluated the zero-shot performance of LLaVA models of varying sizes on radio benchmarks, comparing with alternative open-weight VLMs and a representative commercial model (OpenAI GPT 4.1).
Results are reported in Figure~\ref{fig:eval-zeroshot} and discussed in the following paragraphs. For each benchmark, we report the classification F1-score for individual classes, as well as the average F1-score across all classes (labelled as 'AVG' in the plots).

\subsubsection{Open-weight models}
In Figure~\ref{fig:eval-zeroshot}, we present the benchmark evaluation results for the base LLaVA-OneVision models (0.5B, 7B, 72B), shown in blue histograms, compared against alternative open-weight VLM models: TinyLLaVA 3.1B (green histogram), Qwen2VL models (2B, 7B, 72B) (red histograms), and InternVL models (2B, 8B) (orange histograms).

As expected, smaller models (0.5B-3.1 B) perform consistently worse across most benchmarks, while larger models (Qwen2VL 72B, InternVL 8B, and LLaVA 72B) tend to achieve the best performance, particularly in B3 (radio galaxy detection), B4 (artifact detection), and B6 (FR-I vs. FR-II classification). In B1 (extended/diffuse source detection) and B2 (morphology classification), performance remains generally low across all models, with no significant advantage for any specific one. The best results are observed in B3 and B4, where LLaVA 7B/72B models achieve competitive or slightly better performance compared to recently released VLMs. For instance, in artifact detection (B4), they attain a respectable 50–60\% F1-score in a zero-shot setting. B5 (peculiar/complex morphology classification) and B6 (FR-I vs. FR-II classification) present significant challenges for all models, including the largest ones. Overall, the results indicate poor performance across all benchmarks, underscoring the need for models specialized in astronomical data.

\begin{figure*}[!htb]
\centering%
\subtable[B1]{\includegraphics[scale=0.45]{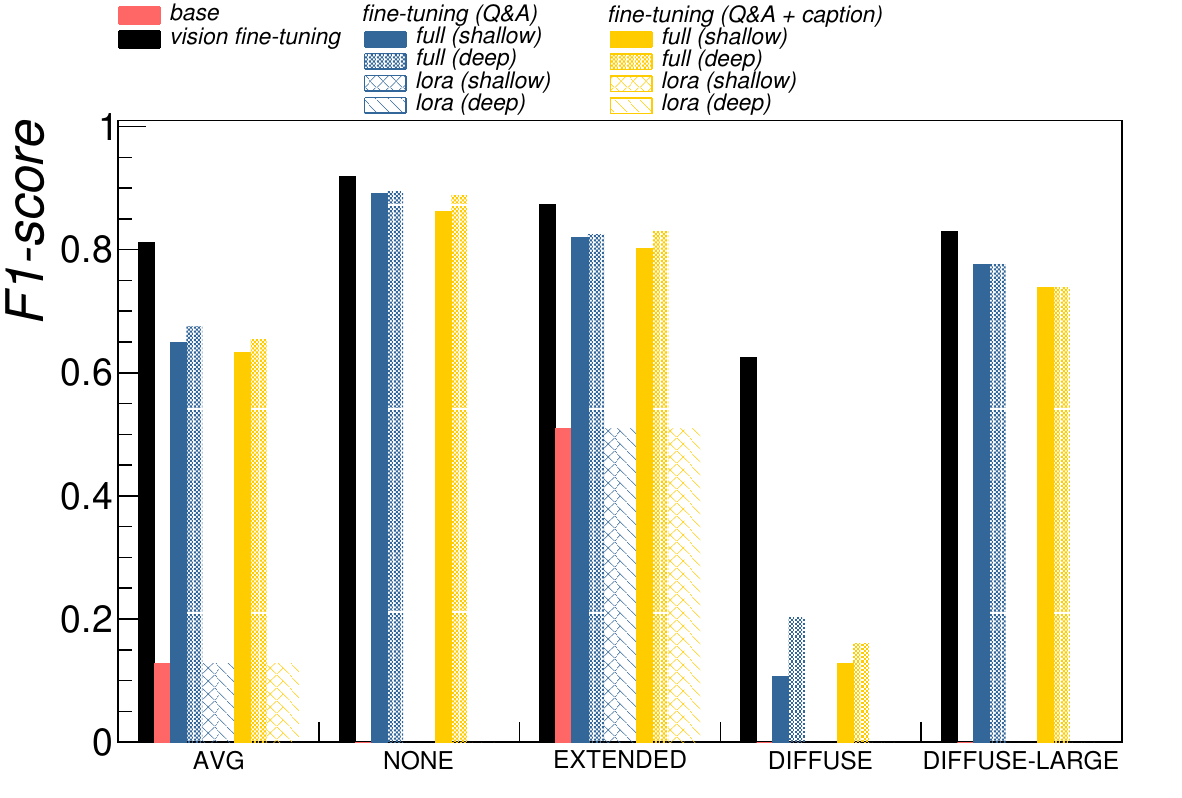}}
\hspace{-0.2cm}%
\subtable[B2]{\includegraphics[scale=0.45]{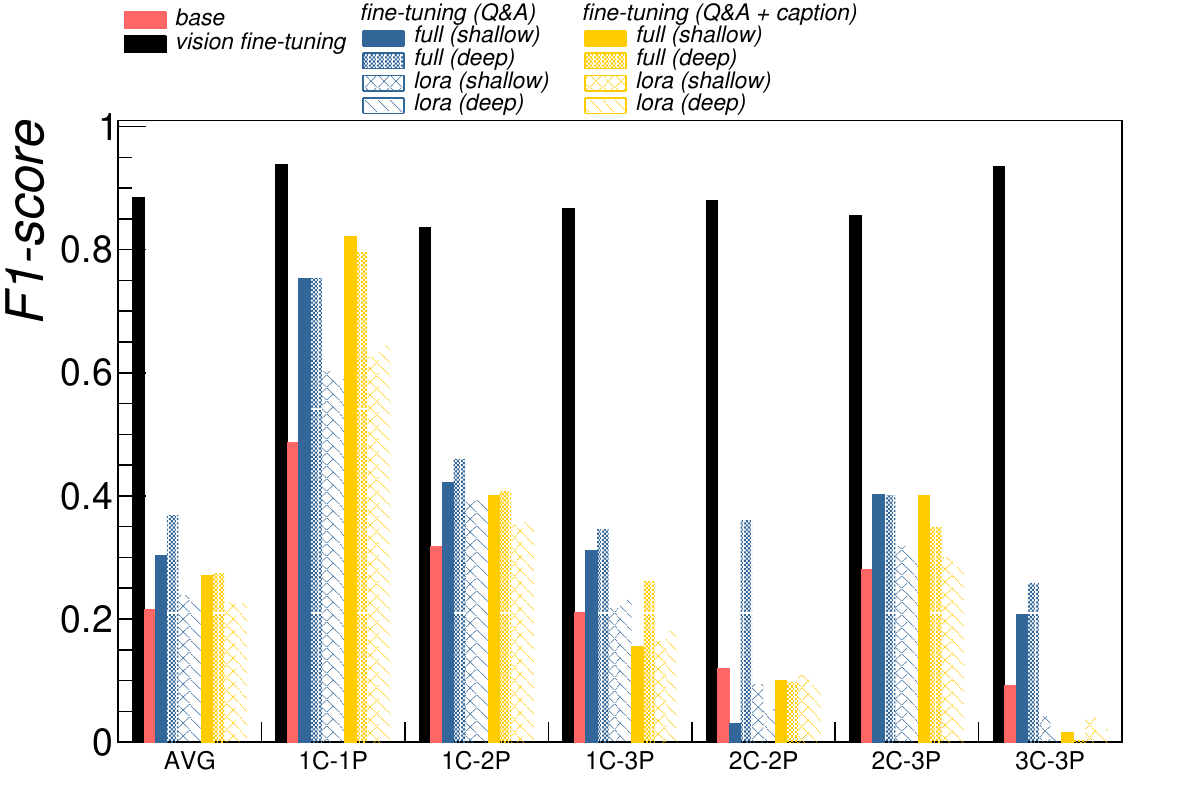}}\\
\vspace{-0.4cm}%
\subtable[B3]{\includegraphics[scale=0.45]{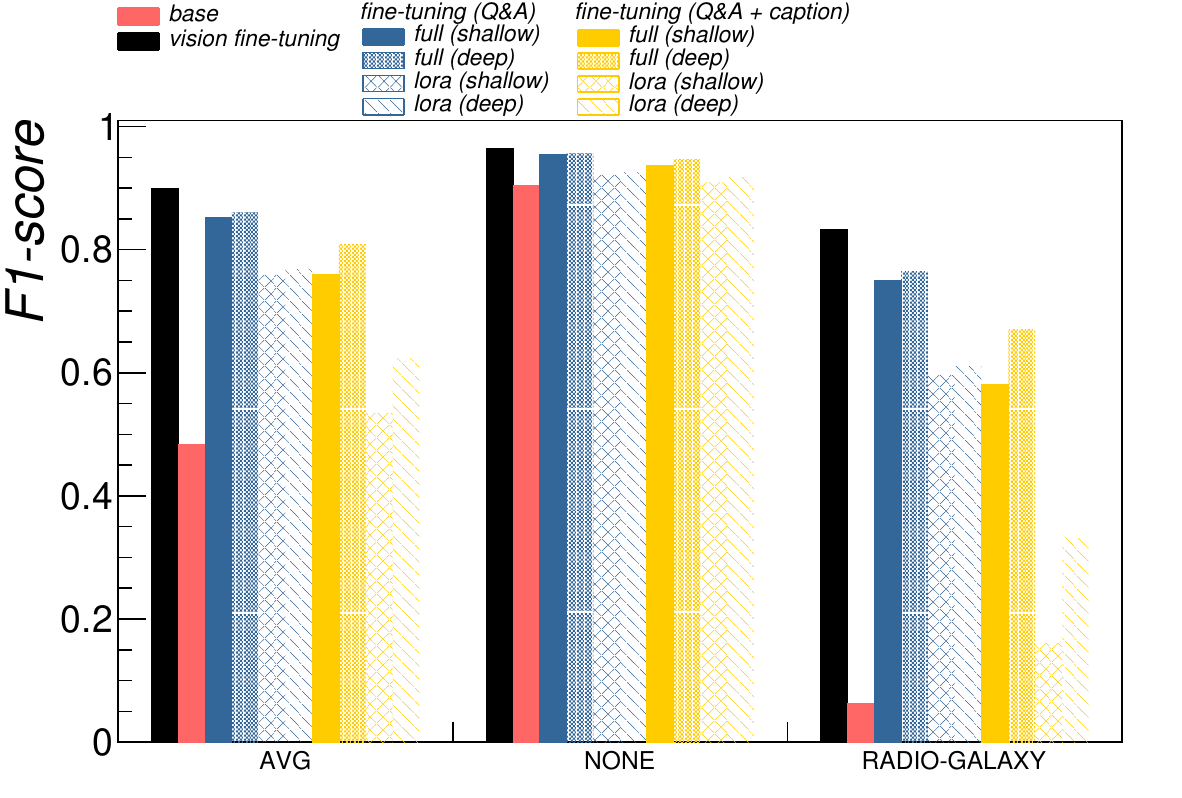}}
\hspace{-0.2cm}%
\subtable[B4]{\includegraphics[scale=0.45]{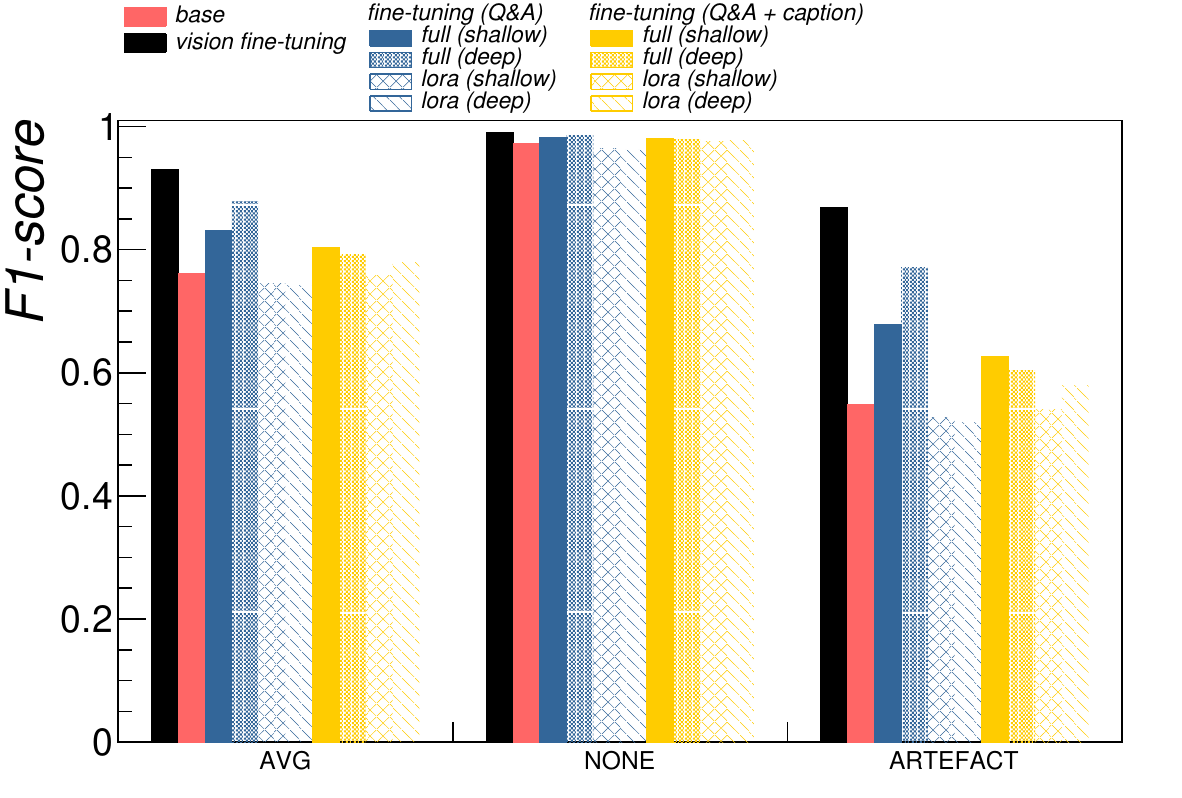}}\\
\vspace{-0.4cm}%
\subtable[B5]{\includegraphics[scale=0.45]{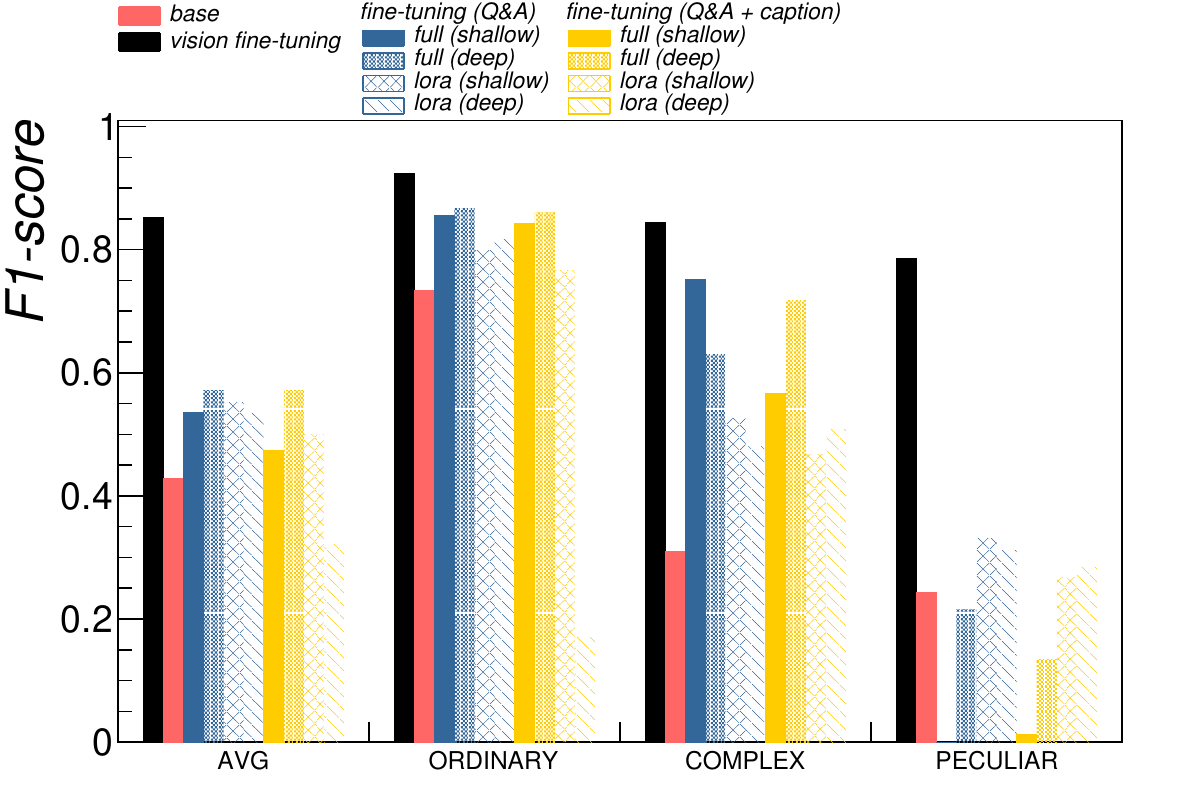}}
\hspace{-0.2cm}%
\subtable[B6]{\includegraphics[scale=0.45]{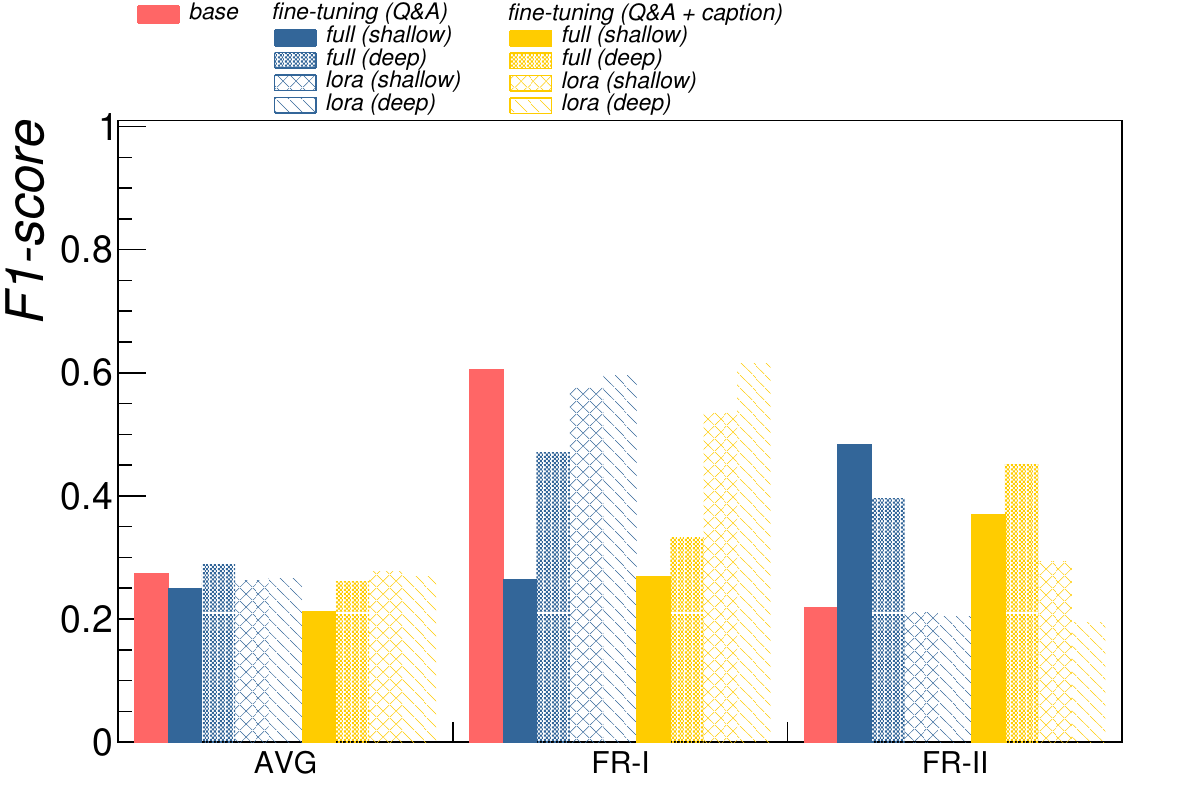}}
\vspace{-0.4cm}%
\caption{Classification F1-scores obtained with the \emph{radio-llava} model on B1–B6 radio benchmarks, comparing fine-tuning on the Q\&A training dataset (blue histograms) and the combined Q\&A and caption datasets (orange histograms). For each training set, results are reported for different training strategies (full vs. LoRA fine-tuning) and training depths (shallow vs. deep). Results from the base model are shown as filled red histograms. Results obtained with a fine-tuned vision-only model (siglip-so400m-patch14-384 encoder) are shown as filled black histograms. The class-averaged F1-scores are labelled as 'AVG'.}%
\label{fig:eval-finetune}
\end{figure*}

\subsubsection{Commercial closed-weight models}
\label{subsec:analysis-zeroshot-commercial}
Performing a comprehensive end-to-end benchmark evaluation across major proprietary solutions (e.g. \texttt{OpenAI GPT}, \texttt{Google Gemini}, \texttt{Anthropic Claude}) is not straightforward, as it would require academic institution to enter into contractual agreements with private providers to cover the cost of executing benchmarks via their APIs. Unlike public user interfaces, these APIs typically operate under separate pricing and access tiers. Nonetheless, we recognise the value of such an analysis for understanding the feasibility and cost-effectiveness of commercial LLM APIs in scientific benchmarking. Therefore, we made an effort to evaluate at least one commercial model - GPT-4.1 via the OpenAI API platform.

Benchmarks were split into multiple sub-tasks of approximately 500 images each to stay within the maximum batch file size limit (200 MB), with each mini-batch consisting of 80K$-$230K input tokens, costing around 0.23\$ - implying a total of $\sim$2.5\$ per benchmark and under 20\$ for the full suite. As Tier 1 users, we were able to run one or two mini-batches per day without exceeding the token-per-day (TPD) limit of 900,000 tokens.

The GPT-4.1 benchmark results, shown in Figure~\ref{fig:eval-zeroshot} as black histograms, indicate superior performance in tasks B1-B3 and B6. In tasks B1 and B6, GPT-4.1 outperforms all open-weight models by a substantial margin $-$ approximately 20\% in average classification score. For tasks B2 and B3, the improvement is more modest, generally below 10\%. Interestingly, GPT-4.1 underperforms in tasks B4 and B5, where its classification metrics fall below those of several open-weight models. 
These results may reflect both the advantage conferred by GPT-4's significantly larger parameter count\footnote{Notably, GPT-4 models are estimated to have approximately 25 times more parameters (around 1.8 trillion parameters from various sources) than the largest open-weight models evaluated in this work} and broader pretraining corpus, as well as limitations in its exposure to domain-specific astronomical concepts or visual patterns. While GPT-4.1 currently achieves the best overall performance in our benchmarks, the relatively small gap in several tasks $-$ combined with the flexibility, transparency, and lower deployment costs of open-weight models $-$ suggests there remains meaningful room for their development and application in specialized astronomical workflows.

\subsection{Fine-tuning performance}
\label{subsec:analysis-finetuning}

\subsubsection{Radio benchmarks}
In Figure~\ref{fig:eval-finetune} we report the classification F1-score of \textit{radio-llava} fine-tuned models obtained on radio benchmarks for each class and overall (labelled as 'AVG'), compared to the base LLaVA-OneVision 7B model (solid red histograms). Blue histograms represent models fine-tuned on the Q\&A dataset, using either deep/shallow full fine-tuning or LoRA fine-tuning. Orange histograms correspond to models fine-tuned on the combined Q\&A and caption datasets. For comparison, we also report baseline metrics (shown as black histograms) obtained using a vision-only classifier that shares the same vision encoder as the LLaVA model (\texttt{siglip-so400m-patch14-384}). This classifier was fine-tuned and evaluated on the same training and test datasets.

With the exception of B6, we observe a general improvement in performance when fully fine-tuning the base model. The performance boost is particularly notable for B1 (extended/diffuse source detection) and B3 (radio galaxy classification), where average classification scores improve by more than 20–30\%. For the remaining tasks, the improvement is more moderate ($\sim$10\%). In contrast, LoRA fine-tuning leads to a clear improvement only in B3 and B5 tasks, with limited gains elsewhere. Deeper fine-tuning results in a modest improvement of only a few percentage points across all tasks, for both full and LoRA fine-tuning strategies. Fine-tuning on caption data (orange histograms) is observed to slightly decrease performance on radio benchmarks. This is somewhat expected, as all radio benchmarks are based on Q\&A tasks rather than descriptive tasks. Caption data, on the other hand, have a positive impact on non-radio benchmarks, as discussed in the next section.

Overall, the achieved metrics remain well below those obtained using a vision-only model specialized for each task, which consistently reaches over 80$-$85\% accuracy across all benchmarks - even after just 10 training epochs. The performance gap is especially notable in task B1, where the vision-only model attains an F1-score of approximately 60\% for diffuse sources $-$ class that multimodal models tend to struggle with, likely due to their underrepresentation in the training set (only 534 images). Similarly, in task B2, the model achieves over 80\% accuracy across all morphological classes. These results also surpass our previous baseline of 74\% average F1-score \citep{RiggiADASS2024}, which was obtained by training a LightGBM classifier on features extracted solely from the SigLIP vision encoder. At present, a fair comparison between the vision-only and LLaVA models cannot be provided for benchmark B6 due to the lack of a shared training dataset. Specifically, the FR-I/FR-II labels used in the LLaVA training were derived from object detection conversations on wide-field images, whereas training a vision-only classifier would require centred cutouts around FR-I/FR-II sources. Previous FR-I/FR-II classification studies \citep{CecconelloPRRS2024, Slijepcevic2024} have achieved over 90\% classification accuracy using smaller fine-tuned encoders (ResNet18/ResNet50). However, those models were trained on an independent subset of the same survey data (MiraBest dataset, VLA survey) used for testing. In contrast, our work trained on different survey data with varying source/image size ratios $-$ specifically, using ASKAP EMU pilot data for training and VLA zoomed-in source images for testing. This difference likely contributes to the poor results observed on B6. We plan to update the dataset accordingly in future work to enable a consistent evaluation across both models.

These findings suggest that the visual encoder provides a strong data representation, even for radio data, justifying our decision to keep it frozen during \textit{radio-llava} fine-tuning. This also indicates that the suboptimal performance of our multimodal models is likely due to visual-language misalignment and the limited size and quality of the training dataset. Indeed, when we attempted to fully fine-tune \textit{radio-llava}, including the vision encoder, we observed only a minor performance improvement ($\sim$2\%). However, it is important to note that, unlike specialized vision encoder models, \textit{radio-llava} was trained to learn multiple radio tasks simultaneously.

\begin{figure*}[!htb]
\centering%
\subtable{\includegraphics[scale=0.45]{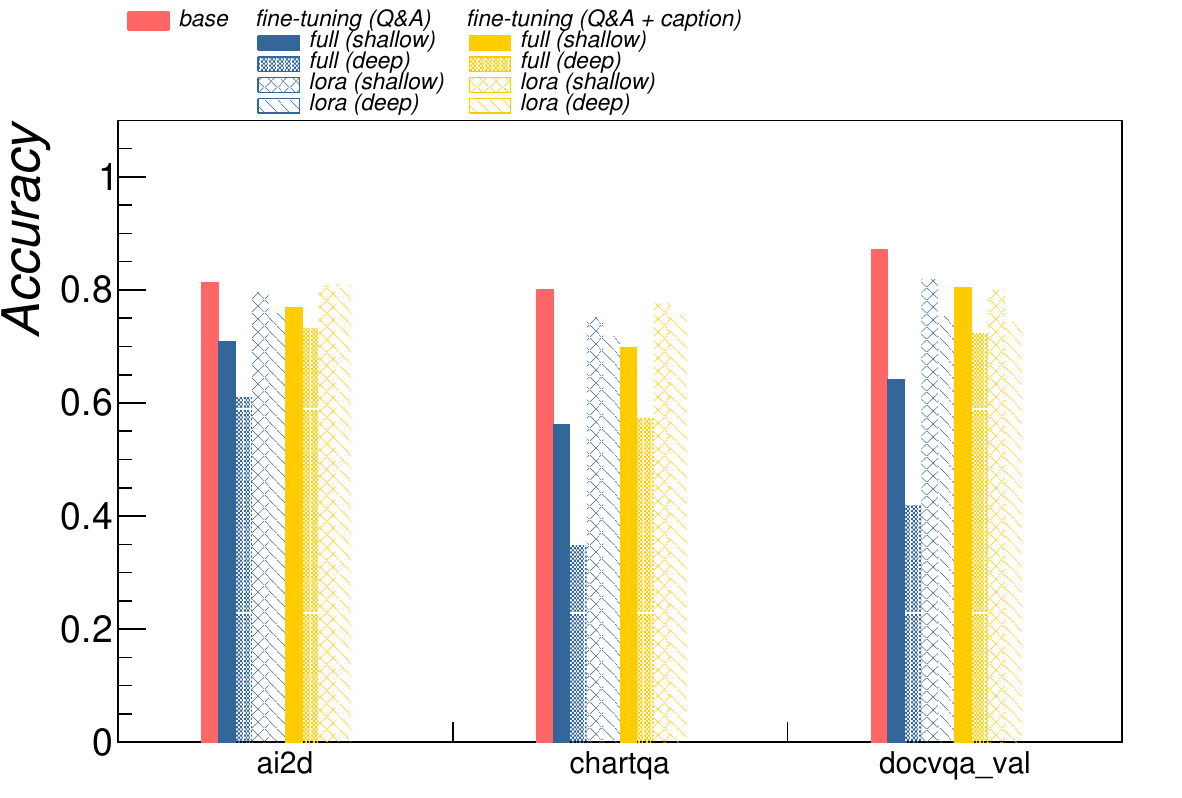}}
\hspace{-0.2cm}%
\subtable{\includegraphics[scale=0.45]{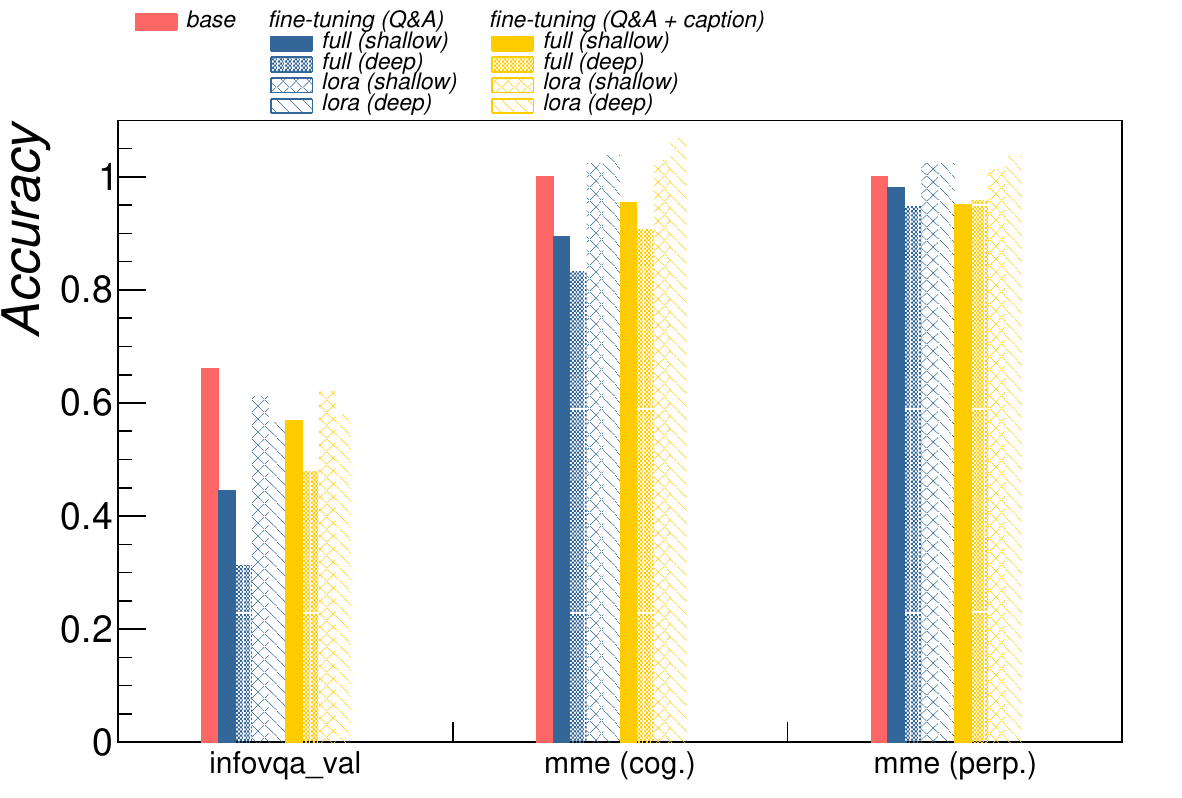}}\\
\vspace{-0.4cm}%
\subtable{\includegraphics[scale=0.45]{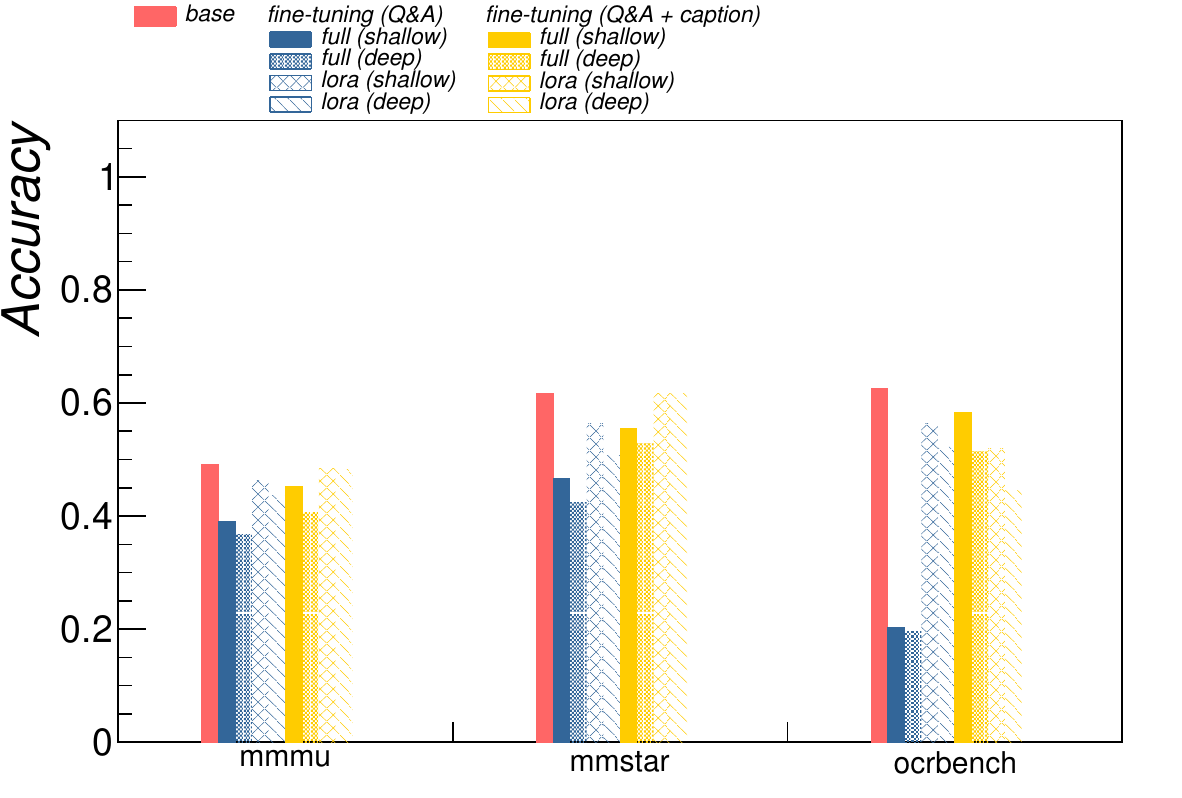}}
\hspace{-0.2cm}%
\subtable{\includegraphics[scale=0.45]{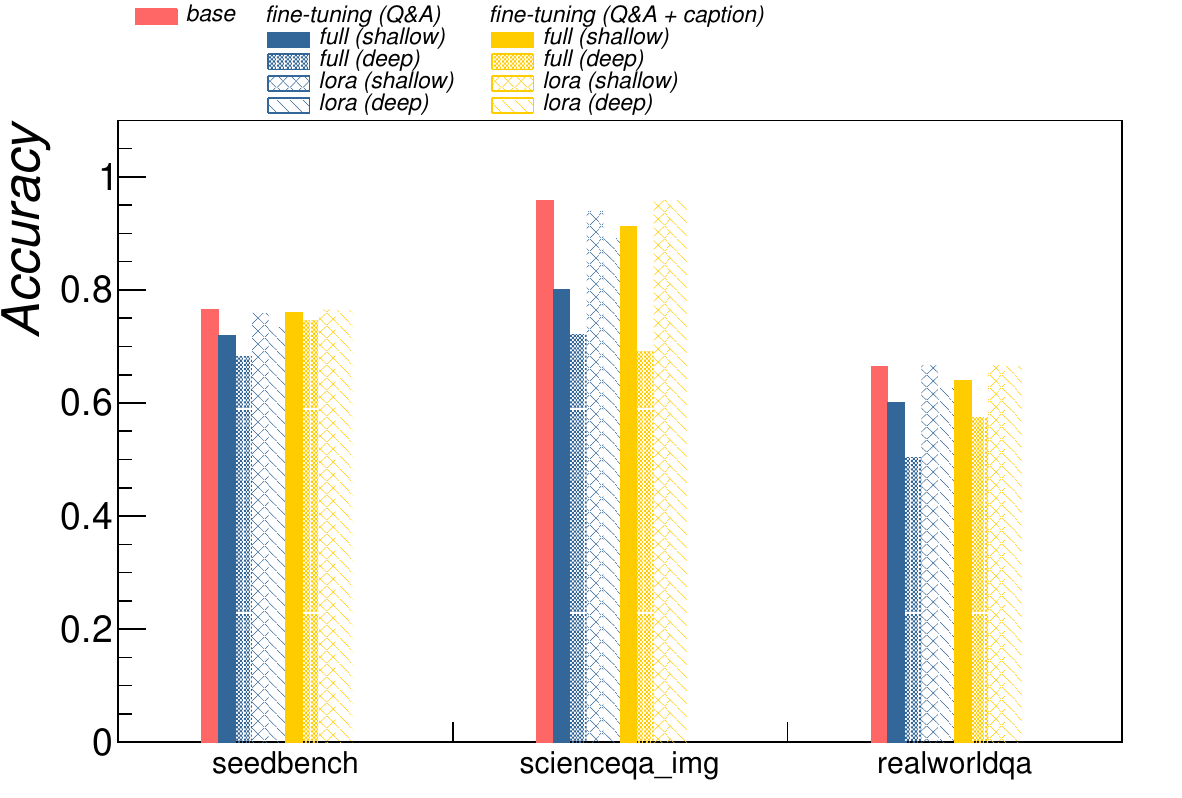}}
\caption{Classification accuracy obtained with the \emph{radio-llava} model on standard non-radio benchmarks (Section~\ref{subsec:evaluation-benchmarks-standard}), comparing fine-tuning on the Q\&A training dataset (blue histograms) and the combined Q\&A and caption datasets (orange histograms). For each training set, results are reported for different training strategies (full vs. LoRA fine-tuning) and training depths (shallow vs. deep). Results from the base model are shown as filled red histograms.}%
\label{fig:eval-finetune-standardbench}
\end{figure*}

\subsubsection{Standard benchmarks}
For comparison, Figure~\ref{fig:eval-finetune-standardbench} reports performance on non-radio benchmarks, using the same colour labelling scheme of Figure~\ref{fig:eval-finetune}. Consistent with previous studies \citep{AstroLLaMA2} specializing LLMs for astronomy, we observed a notable decline in model performance on previously learned tasks compared to the base model (solid red histograms). This task forgetting effect is particularly pronounced in full fine-tuning, becoming significant (more than a 20\% accuracy drop) in deeper training runs. In line with findings by \cite{Biderman2024}, LoRA fine-tuned models achieve lower performance on radio benchmarks but are more robust against task forgetting.

Catastrophic forgetting remains a critical challenge when fine-tuning LLMs. Recent studies \citep{Zhai2023, Zhang2024} have analysed this effect in multimodal models and proposed various strategies to mitigate it. One promising approach, successfully explored for language models by \cite{AstroLLaMA3}, involves expanding and curating the instruction-tuning dataset, followed by merging fine-tuned models with base models, using customizable balancing weights\footnote{The merging tool used in \cite{AstroLLaMA3} $-$ \texttt{MergeKit} \url{https://github.com/arcee-ai/mergekit} $-$ currently supports merging only the LLM components and requires extension to also include the LLaVA adapter and vision components.}.
In this work, we nearly doubled the size of our initial Q\&A dataset, enriching it with more diverse image captions extracted from a large collection of scientific papers. As shown in Figure~\ref{fig:eval-finetune-standardbench}, incorporating caption data (orange histograms) helped recover approximately 10 accuracy points across all standard benchmarks. This confirms trends observed in previous studies and underscores the importance of further curating our training dataset to enable future improvements. 

\subsubsection{Diagnostic analysis}
To assess the impact of the default training configuration on model performance, we fine-tuned the model on the Q\&A dataset with alternative choices of selected hyperparameters, resulting in various model variants, which are labelled and summarised in Table~\ref{tab:model-variants}.
\begin{itemize}
\item Model \texttt{v1} and \texttt{v2} were trained with alternative learning rates (5$\times$10$^{-5}$, 5$\times$10$^{-6}$) compared to the default 10$^{-5}$;
\item Model \texttt{v3} and \texttt{v4} explored alternative learning schedulers: \texttt{v3} used a "faster" warmup phase with \texttt{warmup\_ratio}=0.01 (compared to the 0.03 default), while \texttt{v4} employed a \texttt{cosine\_with\_min\_lr} scheduler with a minimum learning rate of 5$\times$10$^{-6}$, instead of the unconstrained \texttt{cosine} scheduler;
\item Model \texttt{v5} was trained on 32 4 GPU nodes (\texttt{batch\_size}=1, \texttt{gradient\_accumulation\_step}=2) to obtain a larger effective batch size of 256 (compared to the default 8);
\item Model \texttt{v6} used LoRA fine-tuning with larger ranks ($r$=128, alpha=256) instead of the previously tested $r$=64, alpha=128.
\end{itemize}
Additionally, we examined the impact of the user prompt by repeating the benchmark evaluation with a more structured prompt version.

Figure~\ref{fig:finetune-metrics-altpars} in the Appendix compares average metrics across all radio tasks for the original and variant models: the black solid histogram represents the original model, the black dashed histogram represents the original model with the alternative prompt, and the coloured histograms correspond to the model variants. From the results, we conclude that the alternative configurations tested do not lead to performance improvements. Thus, the suboptimal performance is unlikely due to non-optimal hyperparameter selection but rather to dataset quality limitations. Furthermore, the reported metrics show minimal variation with the adopted prompt.

To assess the impact of using automated data curation via InternVL, we initially conducted experiments using a fully templated, uncurated dataset without InternVL-generated variation. While benchmark performance metrics remained broadly similar, we observed a marked degradation in the model outputs, with responses frequently mirroring the rigid structure of the templates. This motivated us to setup an automated data curation to enhance linguistic diversity and model conversational ability.

\begin{table}[htb]
\centering%
\scriptsize%
\caption{Summary of fine-tuned models with alternative hyperparameter configurations.}
\begin{tabular}{ll}
\hline%
\hline%
Model variant &  Parameters\\%
\hline%
\texttt{v1} & lr=5$\times$10$^{-5}$\\%
\texttt{v2} & lr=5$\times$10$^{-6}$\\%
\texttt{v3} & \texttt{warmup\_ratio}=0.01\\%
\texttt{v4} & \texttt{cosine\_with\_min\_lr} (lr=5$\times$10$^{-6}$)\\%
\texttt{v5} & \texttt{effective\_batch\_size}=256\\%
\texttt{v6} & LORA rank=128, alpha=256\\%
\hline%
\hline%
\end{tabular}
\label{tab:model-variants}
\end{table}

\section{Summary}
\label{sec:summary}
In this work, we investigated the feasibility of using small-scale Vision-Language Models (VLMs) as AI assistants for analysing radio images, enabling tasks such as source classification, identification of specific object classes, and data exploration for quality assessment. Unlike conventional deep learning approaches, VLMs offer a more flexible, natural-language-driven interaction, reducing the need for complex coding or task-specific model adaptation. To this end, we fine-tuned LLaVA, a state-of-the-art VLM, on a custom dataset of over 59,000 radio images paired with instruction-based queries, along with an additional 38,000 image-caption pairs extracted from a large corpus of radio astronomical papers. The fine-tuning process leveraged both Q\&A interactions and descriptive captions, enabling the model to handle a variety of radio analysis tasks, including source morphology classification, extended source detection, and artifact identification. The resulting \textit{radio-llava} model was evaluated across six radio-specific benchmarks (B1$-$B6) and compared against baseline VLMs on non-astronomical multimodal tasks. Fine-tuned models and the developed software have been publicly released.

Our key findings can be summarised as follows:
\begin{itemize}
\item \emph{Fine-tuning improves performance}: Compared to the base model, \textit{radio-llava} exhibits significant performance gains on radio benchmarks, particularly in extended source detection (B1) and radio galaxy classification (B3), with F1-score improvements exceeding 20–30\%;
\item \emph{Challenges in multimodal alignment}: Despite fine-tuning, pure vision models still outperform VLMs, suggesting that visual-language alignment remains a limiting factor. Full fine-tuning of both vision and language components resulted in only marginal improvements ($\sim$2\%);
\item \emph{Task forgetting effect}: While fine-tuned models improve in radio-specific tasks, they suffer from catastrophic forgetting when evaluated on general multimodal benchmarks. This effect is more severe for full fine-tuning ($\sim$20\% accuracy drop), while LoRA fine-tuned models exhibit better retention of prior knowledge. Fine-tuned models were also observed to exhibit degraded conversational capabilities;
\item \emph{Impact of caption data}: Incorporating descriptive captions from scientific literature into the training set enhances model generalization, helping recover $\sim$10 accuracy points on standard multimodal benchmarks while also improving instruction-following abilities.
\end{itemize}
These findings highlight the potential of compact multimodal models for radio astronomy while also revealing key limitations that require further research to fully match the performance of specialized vision models. Future efforts should focus on improving vision-language alignment, curating larger, high-quality training datasets, and exploring hybrid fine-tuning strategies also for larger models ($\sim$70B) to mitigate task forgetting while maximizing domain-specific performance. Additionally, we plan to leverage the multi-image processing capabilities of the LLaVA-OneVision model for in-context learning of analysed tasks. Future investigations will also explore its performance on new tasks requiring comparative analysis across multiple images, such as image retrieval of known source classes.



\begin{acknowledgement}
This scientific work uses data obtained from Inyarrimanha Ilgari Bundara / the Murchison Radio-astronomy Observatory. We acknowledge the Wajarri Yamaji People as the Traditional Owners and native title holders of the Observatory site. CSIRO’s ASKAP radio telescope is part of the Australia Telescope National Facility (\url{https://ror.org/05qajvd42}). Operation of ASKAP is funded by the Australian Government with support from the National Collaborative Research Infrastructure Strategy. ASKAP uses the resources of the Pawsey Supercomputing Research Centre. Establishment of ASKAP, Inyarrimanha Ilgari Bundara, the CSIRO Murchison Radio-astronomy Observatory and the Pawsey Supercomputing Research Centre are initiatives of the Australian Government, with support from the Government of Western Australia and the Science and Industry Endowment Fund.

We acknowledge ISCRA for awarding this project access to the LEONARDO supercomputer, owned by the EuroHPC Joint Undertaking, hosted by CINECA (Italy). Additional computing resources for this work were provided by the INAF "CIRASA" (\emph{Collaborative and Integrated platform for Radio Astronomical Source Analysis}) project, the Italian PON 2014-2020 "MOSAICO" project, and the Italian PNRR Project IR0000034 "STILES" (\emph{Strengthening the Italian leadership in ELT and SKA}) project.
\end{acknowledgement}

\paragraph{Funding Statement}
This work received funding from the INAF "SCIARADA" grant.

\paragraph{Competing Interests}
None

\paragraph{Data Availability Statement}
\label{sec:data-availability}
The software code used in this work is publicly available under the GNU General Public License v3.0\footnote{\scriptsize{\url{https://www.gnu.org/licenses/gpl-3.0.html}}} on the GitHub repository \url{https://github.com/SKA-INAF/radio-llava/}. The trained model weights have been made available in this \emph{Hugging Face} repository: \url{https://huggingface.co/inaf-oact-ai}.

\printendnotes

\newpage%
\appendix%
\renewcommand\thefigure{\thesection.\arabic{figure}} 
\renewcommand{\thesection}{\Alph{section}}

\section{Training Datasets}
\setcounter{figure}{0}
\setcounter{table}{0}
\renewcommand{\thetable}{A\arabic{table}}

\label{appendix:training-datasets}
\subsection{Coarse-grained radio datasets}
We describe below the annotated datasets used to create the conversational train dataset.

\subsubsection*{\textbf{radioimg-multilabel dataset}}
\label{subsec:radioimg-multilabel-dataset}
The dataset currently includes a collection of 19,060 annotated radio images taken from multiple radio surveys, carried out both in the Galactic Plane and outside: 
\begin{itemize}
\item SARAO MeerKAT Galactic Plane Survey (SMGPS) \citep{Goedhart2024}: 2,704 images (14.2\%)
\item ASKAP EMU main survey \citep{EMUMainSurvey}: 4,456 images (23.4\%)
\item ASKAP EMU pilot survey \citep{Norris2021}: 5,860 images (30.7\%)
\item ASKAP EMU pilot Galactic Plane surveys \citep{Umana2021}: 6,040 images (31.7\%) 
\end{itemize}
We manually assigned the following labels to each image:
\begin{itemize}
\item \texttt{BACKGROUND}: If the image is purely background noise, e.g. no sources are visible. Typically, this label is set for frames located at the map borders;
\item \texttt{COMPACT}: if point sources or compact sources comparable with the synthesized beam size (say $<$10 times the beam) are present. Double or triple sources with point-like components also fall into this category;
\item \texttt{EXTENDED}: if any extended source is visible, e.g. a compact source with extension $>$10 $\times$ beam;
\item \texttt{RADIO-GALAXY}: if any extended source is visible with a single- or multi-island morphology, suggesting that of a radio galaxy (e.g. core + lobes);
\item \texttt{DIFFUSE}: if any diffuse source is visible, typically having small-scale (e.g. $<$few arcmin) and roundish morphology;
\item \texttt{DIFFUSE-LARGE}: if any large-scale (e.g. covering half of the image) diffuse object with irregular shape is visible;
\item \texttt{FILAMENT}: if any extended filamentary structures is visible;
\item \texttt{ARTIFACT}: if any ring-shaped or ray-like artefact is visible, e.g. typically around bright resolved sources;
\item \texttt{PECULIAR}: if any object is found with peculiar/anomalous morphology;
\item \texttt{MOSAICKING}: if any residual pattern of the mosaicking process used to produce the image is present;
\item \texttt{BORDER}: if the image contains blank/NaN regions along its borders.
\end{itemize}
More than one label can be assigned to each image, depending on the object/features the user recognises in the image. The number of images that have been assigned each specific label is reported in Table~\ref{tab:radioimg-label-counts}.

The dataset was split into two samples. The first sample, containing 13,342 images, was used to generate the user-assistant conversations for the training Q\&A dataset starting from the template image description created from assigned class labels, as described in Section~\ref{subsec:training-datasets-qa}. The rest of the dataset, consisting of 5,718 images, was used to evaluate the performance of trained models.

\begin{table}[htb]
\centering%
\scriptsize%
\caption{The number of images in the \textit{radioimg-multilabel} dataset that have been assigned each specific label. Multiple labels can be assigned to a single image, as they are not mutually exclusive.}
\begin{tabular}{ll}
\hline%
\hline%
label & \#images\\%
\hline%
\texttt{BACKGROUND} & 116\\%
\texttt{COMPACT} & 18,671\\%
\texttt{EXTENDED} & 3,279\\%
\texttt{RADIO-GALAXY} & 3,269\\%
\texttt{DIFFUSE} & 757\\%
\texttt{DIFFUSE-LARGE} & 1438\\%
\texttt{FILAMENT} & 50\\%
\texttt{ARTIFACT} & 1283\\%
\texttt{PECULIAR} & 439\\%
\texttt{MOSAICKING} & 260\\%
\texttt{BORDER} & 453\\%
\hline%
\hline%
\end{tabular}
\label{tab:radioimg-label-counts}
\end{table}

\subsubsection*{\textbf{rgz-smorph dataset}}
The dataset currently includes a collection of 9,570 radio images extracted from the VLA FIRST survey \citep{Becker1995} and annotated in the Radio Galaxy Zoo (RGZ) crowdsourced project \citep{Banfield2015}. Each image is centred and zoomed on radio sources of 6 different morphological classes, defined on the basis of the observed number of components (C) and peaks (P) as follows: \texttt{1C-1P}, \texttt{1C-2P}, \texttt{1C-3P}, \texttt{2C-2P}, \texttt{2C-3P}, \texttt{3C-3P}. The entire dataset was split into two samples. The first one, containing 5,735 images ($\sim$1000 per class), was used to create the conversational dataset, while the remaining sample (3,835 images, $\sim$600 per class) was reserved for model evaluation scopes.

\subsubsection*{\textbf{smgps-extcat dataset}}
The dataset currently includes a collection of 17,062 radio images extracted from the SMGPS survey \citep{Goedhart2024}, each centred and zoomed\footnote{The original image crop size is set to 1.5 times the size of the source bounding box.} on radio sources listed in the SMGPS extended source catalogue \citep{Bordiu2025}. This includes single- or multi-island sources with morphologies classified as: \texttt{EXTENDED} or \texttt{DIFFUSE}. Furthermore, a fraction of the catalogued sources also have an astronomical class label obtained either through morphological considerations or cross-matching with various Galactic source catalogues (see \citealt{Bordiu2025} for details). Available class labels are: \texttt{GALAXY} (radio galaxy), \texttt{HII} (\hii{} region), \texttt{PN} (planetary nebula), \texttt{SNR} (supernova remnant), \texttt{PULSAR} (pulsar), \texttt{STAR} (generic radio star), \texttt{YSO} (young stellar objects), \texttt{LBV} (luminous blue variable star), \texttt{WR} (Wolf-Rayet star), \texttt{HMXB} (high-mass X-ray binary), \texttt{LMXB} (low-mass X-ray binary). Sources cross-matching to multiple catalogues have more than one label assigned.
All the above source annotations are taken into account to generate the conversational dataset.

\subsection{Fine-grained radio datasets}
\label{subsec:fine-grained-datasets}
We describe below the annotated datasets used to create conversational train datasets that contain precise object localization information.

\subsubsection*{\textbf{caesar-mrcnn dataset}}
The dataset currently contains 12,774 annotated radio images taken from different surveys, such as the VLA FIRST \citep{Becker1995}, ATCA Scorpio \citep{Umana2015}, and ASKAP-EMU Scorpio \citep{Umana2021} surveys. The annotation data include bounding boxes, segmentation masks and classification labels for all radio object identified in the images (38,342 objects, including both real and spurious sources). Objects are classified into five possible classes:
\texttt{SPURIOUS}, \texttt{COMPACT}, \texttt{EXTENDED}, \texttt{EXTENDED-MULTISLAND}, \texttt{FLAGGED}. A detailed explanation of the labelling scheme is provided in the reference publication \citep{RiggiMaskRCNN}.
The entire dataset was used to produce the Q\&A training dataset.

\subsubsection*{\textbf{emu-pilot-rgcat dataset}}
The dataset currently contains 10,414 annotated radio images taken from the ASKAP EMU pilot survey \citep{Norris2021}, each containing at least one extended radio source. Annotation data have been extracted from EMU pilot RG-CAT catalogue \citep{Gupta2024}, including bounding boxes and classification labels for radio objects present in the images. Objects in the original catalogue are classified into  six possible radio galaxy morphology classes: 
\begin{itemize}
\item \texttt{C}: compact radio galaxies;
\item \texttt{FR-I}: radio galaxies of Fanaroff-Riley type I;
\item \texttt{FR-II}: radio galaxies of Fanaroff-Riley type II;
\item \texttt{FR-x}: radio galaxies with mixed or hybrid morphology, showing characteristics of both FR-I and FR-II galaxy classes;
\item \texttt{R}: radio galaxies with single-peak resolved morphology;
\item \texttt{Pec}: radio galaxies with a peculiar morphology;
\end{itemize}
A total of 185,294 objects were annotated according to RG-CAT catalogue.

From a visual inspection of the data, we note that various objects classified as compact (\texttt{C}) should be rather considered as belonging to the \texttt{EXTENDED} class in the classification scheme adopted in the \textit{caesar-mrcnn} dataset (see previous paragraph). To make the two fine-grained datasets more comparable, we applied the following processing steps. We first extracted objects from EMU pilot images using \textit{caesar-mrcnn} trained model \citep{RiggiMaskRCNN}. As a result, we obtained a list of detected objects classified with the \textit{caesar-mrcnn} classification scheme, that was cross-matched with the original RG-CAT object collection. 
This was extended and complemented according to the match results. Objects with a match ($\sim$78\%) were also given a \textit{caesar-mrcnn} label. Objects detected by the \textit{caesar-mrcnn} model but missed in the RG-CAT were added to the final collection, including a total of 231,439 objects.
The obtained source annotations were taken into account to generate the conversational dataset.

\section{Image multi-modal benchmarks}
\label{appendix:image-standard-benchmarks}

\subsection*{AI2}
This benchmark\footnote{\url{https://huggingface.co/datasets/lmms-lab/ai2d}} consists of 3,088 image-based Q\&A pairs on annotated grade school science diagrams from the AI2 Diagrams (AI2D) dataset \citep{aid-dataset}.

\subsection*{ChartQA}
This benchmark\footnote{\url{https://huggingface.co/datasets/lmms-lab/ChartQA}} contains 2,500 image-based Q\&A pairs on real-world charts in various formats (pie, bar) from the ChartQA dataset \citep{chartqa-dataset}.

\subsection*{DocVQA}
This benchmark\footnote{\url{https://huggingface.co/datasets/lmms-lab/DocVQA}} contains 16,626 image-based Q\&A pairs on document of various types and content, sourced from the DocVQA dataset \citep{docvqa-dataset}.

\subsection*{InfoVQA}
This benchmark\footnote{\url{https://huggingface.co/datasets/lmms-lab/DocVQA}, see InfographicVQA validation data split} contains 2,801 image-based Q\&A pairs on document infographics of various types and content, sourced from the InfographicVQA dataset \citep{infovqa-dataset}.

\subsection*{MME}
This benchmark\footnote{\url{https://huggingface.co/datasets/lmms-lab/MME}} consists of 2,374 image-based Q\&A pairs from the MME dataset \citep{mme-dataset}, designed to evaluate multimodal models' perception and cognition abilities. Perception tasks include OCR, recognition of coarse-grained objects (e.g., object presence, count, position, and colour) and fine-grained objects (e.g., identification of movie posters, celebrities, scenes, landmarks and artworks). Cognition tasks cover common sense reasoning, numerical calculation, text translation, and code reasoning.

\subsection*{MMMU}
This benchmark\footnote{\url{https://huggingface.co/datasets/lmms-lab/MMMU}, see validation data split} consists of 900 image-based Q\&A pairs from the MMMU dataset \citep{mmmu-dataset}, designed to assess multimodal perception and reasoning abilities across various image formats, including charts, diagrams, maps, tables, music sheets, and chemical structures. The images are sourced from college exams, quizzes, and textbooks spanning six disciplines: Art \& Design, Business, Science, Health \& Medicine, Humanities \& Social Science, and Tech \& Engineering.

\subsection*{MMStar}
This benchmark\footnote{\url{https://huggingface.co/datasets/Lin-Chen/MMStar}} contains 1,500 image-based Q\&A pairs from the MMStar dataset \citep{mmstar-dataset}, designed to evaluate multimodal models across six core capabilities: Coarse Perception, Fine-grained Perception, Instance Reasoning, Logical Reasoning, Science \& Technology, Mathematics.

\subsection*{OCRBench}
This benchmark\footnote{\url{https://huggingface.co/datasets/echo840/OCRBench}} consists of 1,000 image-based Q\&A pairs from the OCRBench dataset \citep{ocr-dataset}, designed to assess Optical Character Recognition (OCR) capabilities across various domains, including multilingual text, handwritten text, non-semantic text, and mathematical expression recognition.

\subsection*{SEED-Bench}
This benchmark\footnote{\url{https://huggingface.co/datasets/lmms-lab/SEED-Bench}} contains 17,990 image-based Q\&A pairs from the SEED-Bench dataset \citep{seedbench-dataset}, designed to evaluate multimodal model capabilities across various domains, including scene understanding, instance identity/attribute/location/counting, spatial relations, instance interaction, visual reasoning, text recognition, action recognition/prediction, procedure understanding.

\subsection*{ScienceQA-IMG}
This benchmark\footnote{\url{https://huggingface.co/datasets/lmms-lab/ScienceQA}, see ScienceQA-IMG test data split} consists of 2,017 image-based Q\&A pairs from the ScienceQA dataset \citep{scienceqa-dataset}, featuring multiple-choice questions spanning diverse topics across three subjects: natural science, social science, and language science.

\subsection*{RealWorldQA}
This benchmark\footnote{\url{https://huggingface.co/datasets/lmms-lab/RealWorldQA}} contains 765 image-based Q\&A pairs from the RealWorldQA dataset, featuring real-world scenarios, including images captured from vehicles.

\onecolumn%

\section{Supplementary plots/tables}
\label{appendix:addon_plots}
\setcounter{figure}{0}
\setcounter{table}{0}
\renewcommand{\thetable}{C\arabic{table}}

\newcommand{\modelbase}{\texttt{llava-ov}}
\newcommand{\modeltuned}{\texttt{radio-llava}}

\begin{table}[!h]
\begin{minipage}{0.99\textwidth}
\centering  
\vspace{-4mm}
\scalebox{0.88}{
\begin{tabular}{l p{8cm} p{8cm} }
\hline%
&  \includegraphics[height=4cm]{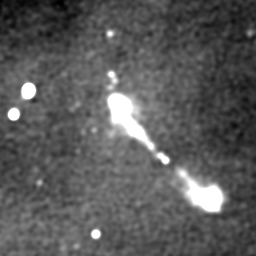} & \includegraphics[height=4cm]{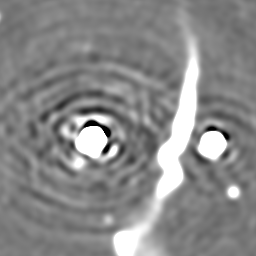}\\
\toprule%
User & \multicolumn{2}{l}{You are given a radio astronomical image as input. Answer to the question below, strictly following the provided instructions.} \\%
\vspace{-0.4cm}%
& \multicolumn{2}{l}{}\\%
& \multicolumn{2}{l}{\#\# Question: Can you provide a brief description of the image?} \\%
& \multicolumn{2}{l}{\#\# Instructions:} \\%
& \multicolumn{2}{l}{$-$ \scriptsize{\textit{Base your description solely on the objects and patterns visible in the image.}}}\\%
& \multicolumn{2}{l}{$-$ \scriptsize{\textit{Do not include details that are not present in the image.}}} \\%
& \multicolumn{2}{l}{$-$ \scriptsize{\textit{Use a scientific and astronomical writing style.}}} \\%
& \multicolumn{2}{l}{$-$ \scriptsize{\textit{Limit your response to a maximum of four sentences.}}} \\%
\cline{2-3}%
\modelbase{} & The image depicts a section of the sky with several bright, white spots against a dark background. These spots are likely to be celestial objects such as stars or galaxies. The distribution and brightness of these objects suggest they could be part of a larger astronomical structure, possibly a galaxy cluster or a group of stars within a constellation. & The image depicts a radio astronomical observation, likely of a celestial object such as a galaxy or a star system. It shows two bright, circular regions that could represent the centers of these objects, surrounded by fainter, ring-like structures that might be indicative of surrounding material or disk-like formations. The central bright spots are possibly the cores of the objects, emitting intense radio signals. The faint lines and patterns suggest variations in intensity or different emission mechanisms within the observed region. \\%
\cline{2-3}%
\blue{\modeltuned{} (Q\&A)} & \blue{The image is a radio astronomical image cutout extracted from a larger radio-continuum Stokes-I map produced by an interferometer telescope, centred and zoomed on a multi-island extended radio source, consisting of 3 islands.} & \blue{The image is a radio astronomical cutout featuring various point-like and compact radio sources, as well as extended radio sources, with some sources exhibiting poor imaging and ring-patterned artefacts.} \\%
\cline{2-3}%
\red{\modeltuned{} (Q\&A + capt.)} & \red{The image shows a radio astronomical cutout of an extended radio galaxy candidate, characterised by its elongated structure and multiple bright spots.} & \red{The image shows a radio astronomical cutout with two prominent sources. One source appears as a bright, compact object near the center of the image, while the other is a more extended source located towards the right side. The background exhibits a complex pattern of concentric rings, likely resulting from the imaging process.} \\%
\midrule
User & \multicolumn{2}{l}{\#\# Question: Do you see any likely radio galaxy with an extended morphology in the image?} \\%
& \multicolumn{2}{l}{\#\# Context: }\\%
& \multicolumn{2}{p{\dimexpr8cm+8cm+2\tabcolsep}}{$-$ \scriptsize{\textit{RADIO GALAXY: a type of active galaxy that emits an exceptionally large amount of radio waves, often extending beyond its visible structure. These galaxies host an active galactic nucleus (AGN), powered by a supermassive black hole (SMBH) at their center, which fuels the production of powerful radio jets and lobes}}.} \\%
& \multicolumn{2}{l}{\#\# Instructions: }\\%
& \multicolumn{2}{l}{$-$ \scriptsize{\textit{Answer the question taking into account the provided context}}.} \\%
\cline{2-3}%
\modelbase{} & Yes, there is a likely radio galaxy with an extended morphology in the image. & Yes, there is a likely radio galaxy with an extended morphology in the image. \\%
\cline{2-3}%
\blue{\modeltuned{} (Q\&A)} & \blue{Yes, the image suggests the presence of an extended radio galaxy candidate due to its elongated and extended morphology.} & \blue{Yes, the image contains candidate radio galaxies with extended morphology.} \\%
\cline{2-3}%
\red{\modeltuned{} (Q\&A + capt.)} & \red{Yes, the image shows characteristics of a radio galaxy with an extended morphology.} & \red{No, the image does not show characteristics of a radio galaxy with an extended morphology.} \\%
\midrule
User & \multicolumn{2}{l}{\#\# Question: Do you see any imaging artifact in the provided image?} \\%
& \multicolumn{2}{l}{\#\# Context: }\\%
& \multicolumn{2}{p{\dimexpr8cm+8cm+2\tabcolsep}}{$-$ \scriptsize{\textit{ARTEFACT: noise pattern with a ring-like or elongated morphology, typically found around bright compact sources in radio images. They are often mistaken for real radio sources but are actually spurious. These patterns arise from imperfections in the imaging processing stage of radio data.}}.} \\%
& \multicolumn{2}{l}{\#\# Instructions: }\\%
& \multicolumn{2}{l}{$-$ \scriptsize{\textit{Answer the question taking into account the provided context}}.} \\%
\cline{2-3}%
\modelbase{} & No & No \\%
\cline{2-3}%
\blue{\modeltuned{} (Q\&A)} & \blue{No, the image does not show any imaging artefacts.} & \blue{Yes, the image contains imaging artefacts.} \\%
\cline{2-3}%
\red{\modeltuned{} (Q\&A + capt.)} & \red{No, there are no imaging artefacts present in the image.} & \red{Yes, there are imaging artefacts present in the image.} \\%
\midrule
\hline%
\hline%
\end{tabular}
}%
\vspace{1mm}
\captionof{table}{User-assistant conversations on sample radio images for base and fine-tuned LLaVA-OneVision models. }
\vspace{-5mm}
\label{tab:llava-conversation-example}  
\end{minipage}
\end{table}

\begin{figure*}[!htb]
\centering%
\includegraphics[scale=1.]{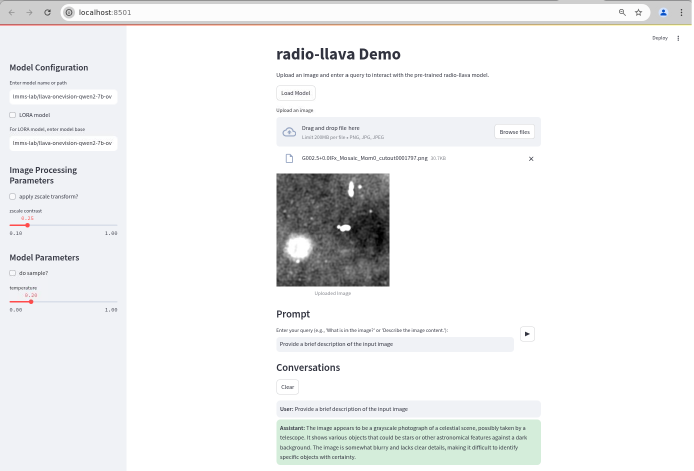}
\caption{A screenshot displaying the Streamlit web application developed for \emph{radio-llava} demo purposes.}%
\label{fig:streamlit-app}
\end{figure*}

\begin{figure*}[!htb]
\centering%
\includegraphics[scale=0.7]{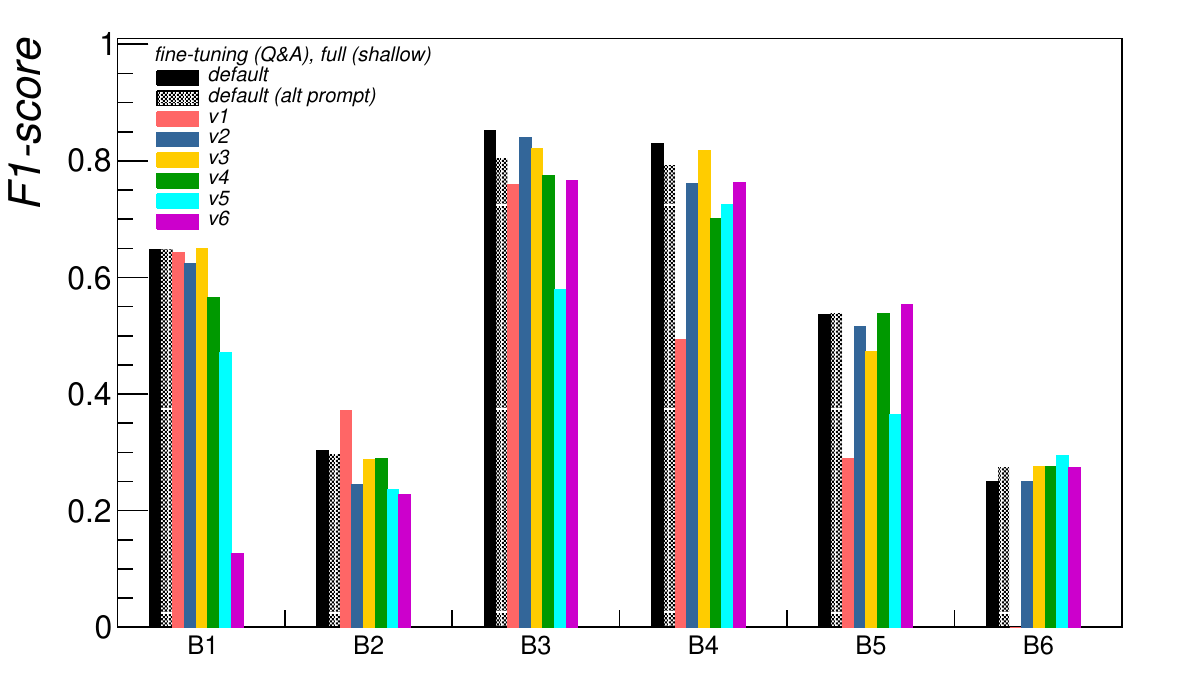}
\caption{Class-averaged classification F1-scores obtained with the \textit{radio-llava} model on B1–B6 radio benchmarks, comparing fine-tuning on the Q\&A training dataset (black solid histograms, labelled as "default") with model variants (v1-v6, coloured histograms), fine-tuned on the same dataset using alternative parameters (see text). The dashed black histogram represents the standard model evaluated on B1-B6 benchmarks using an alternative prompt.}%
\label{fig:finetune-metrics-altpars}
\end{figure*}

\end{document}